\documentclass[lettersize,journal]{IEEEtran}

    \usepackage[pdftex]{graphicx}
    \DeclareGraphicsExtensions{.pdf,.jpeg,.png}
\usepackage{gensymb}
\usepackage{epstopdf}
\usepackage{graphicx}
\usepackage{amssymb}
\usepackage{amsmath}
\usepackage{float}
\usepackage{amsfonts, bm}
\usepackage{color}
\usepackage[noadjust]{cite}
\usepackage{url}
\usepackage{paralist}
\newcommand{\Leb}{\mathbb{L}}

\newcommand{\Reals}{\mathbb{R}}

\newcommand{\mapp}[1]{\bm{#1}}

\DeclareMathOperator{\Ex}{\mathcal{E}}

\DeclareMathOperator{\trace}{tr}

\renewcommand{\theenumi}{\alph{enumi}}

\newtheorem{assumption}{Assumption}
\newtheorem{theorem}{Theorem}

\newtheorem{corollary}{Corollary}
\newtheorem{definition}{Definition}

\title{ 
Large-Scale Self-Powered Vibration Control: Theory and Experiment
}

\author{Connor Ligeikis, Heath Hofmann, and Jeff Scruggs % <-this % stops a space
\thanks{This work was supported by NSF under Grant 2206018. The first author was also supported by an NSF Graduate Research Fellowship. This funding is gratefully acknowledged. Views expressed in this paper are those of the authors and do not necessarily reflect those of the National Science Foundation.}% <-this % stops a space
\thanks{C. Ligeikis is with the Department of Mechanical Engineering, Lafayette College, Easton, PA, 18042. (email: \mbox{ligeikic@lafayette.edu})}
\thanks{H. Hofmann is with the Department of Electrical Engineering \& Computer Science, University of Michigan, Ann Arbor, MI, 48109. (email: \mbox{hofmann@umich.edu})}
\thanks{J. Scruggs is with the Department of Civil \& Environmental Engineering, University of Michigan, Ann Arbor, MI, 48109.
      	(email: \mbox{jscruggs@umich.edu}). }}

\begin{document}

\maketitle

\begin{abstract}   
A self-powered system is a control technology that powers itself by harvesting energy from exogenous disturbances. 
This article details the design and experimental validation of a prototype self-powered vibration control system, for larger-scale applications (i.e., power flows above 1W and forces on the order of 1kN.)
The prototype consists of a linear ballscrew coupled with a permanent-magnet synchronous machine. A custom three-phase inverter is used to control power flow, and a custom half-bridge DC-DC power converter is used to facilitate power flow to and from a storage capacitor. Due to parasitics in the control hardware, feedback laws for self-powered systems must adhere to a feasibility condition tighter than mere passivity. 
This article implements a tractable control design approach that accounts for this feasibility constraint. The control design is validated via hardware-in-the-loop experiments pertaining to a stochastically-excited tuned vibration absorber. 
\end{abstract}


\section{Introduction}

\textit{Self-powered systems} are control technologies that power their operation by harvesting, storing, and reusing energy injected into the plant by exogenous disturbances. A self-powered control system for a mechanical plant can be implemented using an electromechanical transducer embedded within plant and connected to an energy storage subsystem (e.g., a supercapacitor, battery, or mechanical flywheel).
The transducer must be capable of absorbing energy from the plant, and also of injecting energy into it.
Absorbed energy is stored locally, and can be re-injected at a future time to improve dynamic performance. More generally, a self-powered control technology can employ a network of transducers connected to a centralized storage subsystem. In this case, in addition to storing energy for later use, a self-powered system can simultaneously remove energy from one location in the plant and re-injected at another location. Self-powered systems have great potential in a variety of applications for which energy-autonomy is desired or external power supplies are unreliable. 

Self-powered control for vibration suppression applications has been investigated both analytically and experimentally by many researchers (see e.g., \cite{nakano2003self,
nakano2004combined,
khoshnoud2015energy,
choi2009self,
tang2011self,
asai2016nonlinear,li2022self,cheng2024novel}).
These technologies have also been described as ``regenerative'' (see e.g.,
\cite{
jolly1997regenerative,
jolly1997assessing,
margolis2005energy,clemen2014model,
anubi2015energy,
clemen2016regenerative,
liu2013regenerative,
shen2018energy})
or ``energy recycling'' (see e.g.,
\cite{onoda2003energy,
onoda2008performance}). These studies vary significantly in their modeling assumptions and control design techniques. For example, in the papers on regenerative control, it is only the time-average power absorbed by the transducers that is constrained. Also, very few of these studies explicitly account for parasitic losses in the control hardware. 

However, parasitics are unavoidable, and arise in the transducers, power-electronic circuitry, and energy storage subsystems.
As a consequence, self-powered feedback laws must adhere to feasibility conditions tighter than classical feedback passivity. In essence, it must be guaranteed that the transducers absorb sufficient energy to overcome the parasitics. Recently, the first and third authors derived explicit sufficient feasibility conditions which account for parasitics \cite{ligeikis2023feasibility}, and  proposed various control design methodologies that guarantee satisfaction of these conditions \cite{ligeikis2021nonlinear,ligeikis2021feasibility,ligeikis2022lqg}. 

This article details the hardware design, control design, and experimental validation of a prototype self-powered vibration control system. The primary contributions are:
\begin{inparaenum}
    \item A systematic design procedure for self-powered  control hardware comprised of a linear ballscrew actuator coupled with a permanent-magnet synchronous machine (PMSM), a three-phase inverter, a half-bridge DC-DC converter, and a storage capacitor. 
    \item An energy-management control scheme that ensures transducer current control is continuously feasible, and that facilitates bidirectional power flows to and from storage.
    \item A methodology for identifying a parasitic loss model for the prototype, which is subsequently used for control design.
    \item Experimental validation of self-powered feedback control laws designed via the synthesis methodologies proposed in \cite{ligeikis2022lqg}. 
\end{inparaenum}

The rest of this article is organized as follows. Section \ref{sec:review} presents the modeling assumptions and reviews the sufficient feasibility condition for self-powered feedback laws. 
Section \ref{sec:proto} details the design of the prototype hardware. 
Section \ref{sec:rths_example_system} presents the dynamic model of the vibratory system used as the plant in the HiL experiments. 
Section \ref{sec:rths_example_control_design} presents both linear and nonlinear control design techniques for self-powered systems. 
Section \ref{sec:testbed} describes the experimental setup used to conduct the HiL experiments. 
Section \ref{sec:rths_results} presents the experimental results, and Section \ref{sec:conclusion} concludes the article.

\section{Review of Self-Powered Control Theory}\label{sec:review}
\subsection{Modeling and assumptions}

\begin{figure}
	\centering
	\includegraphics[scale=0.85]{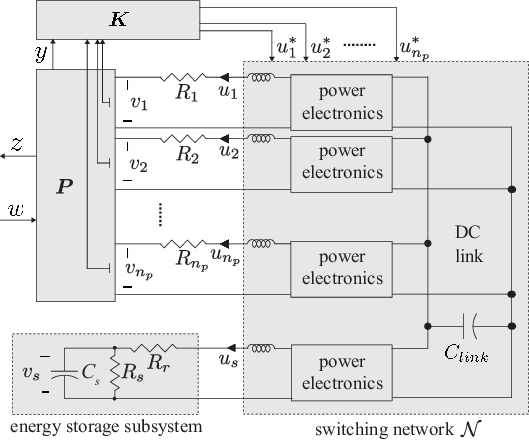}
	\caption{Generic self-powered control system}
	\label{fig:spsa_schematic}
\end{figure}

Figure~\ref{fig:spsa_schematic} shows a schematic of a generic self-powered control system. The exogenous disturbance $w \in \Reals^{n_w}$ injects energy into plant $\mapp{P}$. There are $n_p$ control transducers embedded within the plant at different ports. The dynamics of port $k \in \{1,...,n_p\}$ is characterized by a back-EMF voltage $v_k$, and a colocated current $u_k$. 
Each transducer is connected to a power-electronic drive that regulates current $u_k$ to track a commanded current $u_k^*$ using high-frequency, pulse-width modulation (PWM) switching. The tracking feedback controllers for the drives are designed to provide sufficiently high bandwidth such that $|u_k - u_k^*|$ may be deemed negligible. The command $u_k^*$ is formulated via feedback controller $\mapp{K}$, which accepts measured outputs $y\in\mathbb{R}^{n_y}$ from the plant. Additional plant outputs $z\in\mathbb{R}^{n_z}$ are used to assess closed-loop system performance. 
The power-electronic drives interface each of the $n_p$ ports with a common power bus, called a \emph{DC link}. 
An energy-storage subsystem, depicted as capacitor $C_s$, is also interfaced to the DC link through another power-electronic converter. The power-electronic drives together with the DC link comprise the switching network $\mathcal{N}$ labeled in Figure \ref{fig:spsa_schematic}. This switching network enables bidirectional power flow between the transducers and energy-storage subsystem.
The energy-storage subsystem is assumed to be non-ideal, in the sense that it dissipates energy whenever it transmits power, and because it exhibits leakage. 
These effects are represented by resistances $R_r$ and $R_s$, respectively. 

Regarding $\mathcal{N}$, we make three simplifying assumptions:  

\begin{assumption}\label{lossless_assumption}
$\mathcal{N}$ is lossless.  
\end{assumption}

\begin{assumption}\label{tracking_assumption}
$\mathcal{N}$ facilitates instantaneous tracking between $u$ and $u^*$, assuming $u^*$ is feasible. 
\end{assumption}

\begin{assumption}\label{instantaneous_assumption}
The energy stored in the inductors and capacitor in $\mathcal{N}$ is negligible.
\end{assumption}

Note that Assumption \ref{lossless_assumption} \emph{does not} require the actual power-electronic circuitry to be lossless. This is because parasitic losses in the switching circuits may be subsumed into resistances $R_1, . . ., R_{np}$, as well as $R_r$ and $R_s$. Therefore, these resistances can approximately account for the dissipation of the power electronics, in addition to the actuators and energy storage subsystem. However, it is presumed that the DC link capacitor exhibits negligible leakage dissipation. The total parasitic dissipation, denoted $P_d$, is thus
\begin{equation} \label{eq0_1}
	P_d = u^TRu + R_r u_s^2 + \tfrac{1}{R_s}v_s^2
\end{equation}
where $R \triangleq \text{diag}\left\{R_1,...,R_{n_p}\right\}$. The energy $E_s$ stored in the capacitor $C_s$ is
\begin{equation}
    E_s = \tfrac{1}{2}C_s v_s^2
\end{equation}
Under the above assumptions, it was shown in \cite{ligeikis2023feasibility} that $E_s$ evolves according to 
\begin{equation} \label{Econsv2}
	\tfrac{d}{dt} E_s = -\left( \frac{2}{\tau_s} + \frac{1}{\tau_r} \right)E_s  
	+ \sqrt{ \left(\frac{1}{\tau_r}E_s\right)^2 - \frac{2}{\tau_r} E_s P_e } 
\end{equation}
where 
\begin{equation}\label{Pe1}
    P_e \triangleq u^Tv + u^TR u
\end{equation}
is the power delivered to plant $\mapp{P}$ and where we have defined the time constants $\tau_r = R_rC_s$ and $\tau_s = R_sC_s$. The time constants $\tau_r$ and $\tau_s$ are directly related to the storage subsystem's efficiency. Specifically, a large $\tau_r$ implies poor efficiency of energy transmission to and from storage, and a small $\tau_s$ corresponds to significant leakage of stored energy. 
Note that in order for $E_s(t)\in\Reals_{>0}$ for all $t\in\Reals_{\geqslant 0}$, it must be the case that the argument in the square root in \eqref{Econsv2} must be positive, requiring that $P_e \leqslant \tfrac{1}{2\tau_r} E_s$.

\begin{definition}
Given $\{R,\tau_s,\tau_r\}$ and $v \in \Leb_{2e}^+$, define the set $\mathbb{U}_{SP}(v; R,\tau_s,\tau_r)$ as the set of all $u\in\Leb_{2e}^+$, for which \eqref{Econsv2} has a unique solution satisfying $E_s(t)\in\Reals_{>0}$ for all $t>0$, and for all $E_s(0)\in\Reals_{>0}$.
\end{definition}

\subsection{Self-powered synthetic admittances} 

Consider the case in which there are no supplemental outputs $y$, and the controller $\mapp{K}$ constitutes a colocated mapping $v \mapsto u$. We denote such a colocated controller as $\mapp{Y}$, i.e., 
\begin{equation}
u(t) = -\left( \mapp{Y} v \right) (t) 
\end{equation}
where $\mapp{Y}$ is called a synthetic admittance. If $\mapp{Y}$ produces $u \in \mathbb{U}_{SP}(v;R,\tau_s,\tau_r)$ for all $ v\in \Leb_{2e}^+$, then $\mapp{Y} $ is called a \emph{self-powered synthetic admittance} (SPSA). 
\begin{definition} \label{spsa_def}
Let $\mathbb{Y}_{SP}(R,\tau_s,\tau_r)$ be the set of all SPSAs.
\end{definition}

Next, we briefly review sufficient conditions that guarantee self-powered feasibility. Proofs of the theorems in this section can be found in \cite{ligeikis2023feasibility}. In Section \ref{sec:rths_example_control_design}, we make use of these sufficient conditions to design feasible feedback laws for our prototype self-powered system.

We restrict our attention to linear, finite-dimensional SPSAs which can be represented in state-space form as
\begin{equation}
\mapp{Y} : \left\{ \begin{array}{rl}
    \tfrac{d}{dt} x_Y =& A_Y(t) x_Y + B_Y(t) v \\
    -u =& C_Y(t) x_Y + D_Y(t) v
    \end{array} \right.
    \label{Y_state_space}
\end{equation}
\begin{theorem} \label{LMI_theorem}
Let $\mapp{Y}$ have realization \eqref{Y_state_space}. Then for $R\succ0,~\tau_s>0,$ and $\tau_r \in (0,\infty)$, $\mapp{Y} \in \mathbb{Y}_{SP}(R,\tau_s,\tau_r)$, if there exist a constant matrix $P=P^T\succ 0$ and time-varying matrix $X(t)\in\Reals^{n\times n}$ such that for all $t \in \mathbb{R}_{>0}$,
\begin{align}
    & A_Y^T(t) P + P A_Y(t) + 2\tau_s^{-1} P +X(t) + X^T(t) \preceq 0 \label{lmi_1} \\ & \begin{bmatrix}
        -X(t)-X^T(t) & P B_Y(t) & C_Y^T(t) & -X^T(t) \\ & -\tfrac{1}{2}R^{-1} & D_Y^T(t)-\tfrac{1}{2}R^{-1} & B_Y^T(t) P \\ & & -\tfrac{1}{2}R^{-1} & 0 \\ \mathrm{(sym)}
        & & & -\tfrac{1}{2}\tau_r^{-1} P 
    \end{bmatrix} \nonumber \\ & \hspace{2.7in} \preceq 0
    \label{lmi_2}
\end{align}
\end{theorem}

\begin{definition}
$\mathbb{Y}_{1}(R,\tau_s,\tau_r)$ is the set of all $\mapp{Y} \in \mathbb{Y}_{SP}(R,\tau_s,\tau_r)$, for which the conditions in Theorem \ref{LMI_theorem} hold.
\end{definition}

In the case where $\mapp{Y}$ is LTI we have the following corollary to Theorem \ref{LMI_theorem}.

\begin{corollary}\label{lti_main_theorem}
Let \eqref{Y_state_space} be a realization for $\mapp{Y}$, with $A_Y$, $B_Y$, $C_Y$, and $D_Y$ constant.
Then $\mapp{Y}\in\mathbb{Y}_1(R,\tau_s,\tau_r)$ if there exist time-invariant matrices $P=P^T \succ 0$ and $X\in\Reals^{n\times n}$ such that \eqref{lmi_1} and \eqref{lmi_2} hold. 
\end{corollary}

\begin{figure}
\centering
\includegraphics[scale=1.0,trim=0 0 0 0]{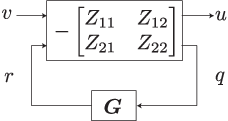}
\caption{LFT representation of SPSA}
\label{fig:LFT}
\end{figure}

We can equivalently formulate $\mapp{Y}$ as the linear fractional transformation (LFT) shown in Figure \ref{fig:LFT}, where $Z(t)$ is a time-varying gain matrix and $\mapp{G}$ is a linear, time-invariant, finite-dimensional system with state-space realization
\begin{equation}
\mapp{G} : \left\{ \begin{array}{rl} 
    \tfrac{d}{dt} x_G =& A_G x_G + B_G q \\
    r =& C_G x_G
    \end{array} \right.
    \label{G_equation}
\end{equation}
where $x_Y=x_G$ and the corresponding $\mapp{Y}$ in \eqref{Y_state_space} is related to $\mapp{G}$ and $Z(t)$ as follows
\begin{align} 
    A_Y(t) =&~ A_G - B_G Z_{22}(t) C_G \label{Y_state_A} \\ 
    B_Y(t) =&~ -B_G Z_{21}(t) \label{Y_state_B} \\
    C_Y(t) =&~ Z_{12}(t)C_G \label{Y_state_C} \\ 
    D_Y(t) =&~ Z_{11}(t) \label{Y_state_D}
\end{align}
\begin{theorem} \label{general_case}
For $R \succ 0,~\tau_s>0,$ and $\tau_r \in (0,\infty)$, $\mapp{Y} \in \mathbb{Y}_{SP}(R,\tau_s,\tau_r)$ if it has an equivalent representation as in Figure \ref{fig:LFT}, in which system $\mapp{G}$ is characterized by \eqref{G_equation}, and if there exists a matrix $P=P^T\succ 0$ such that for all $t\in\Reals_{\geqslant 0}$
\begin{align}
    Z(t)+Z^T(t)-2Z^T(t) W Z(t) \succeq 0& \label{Z_condition} \\
    \tilde{A}_G^T P + P \tilde{A}_G \preceq 0& \label{ABC_cond_gen} \\
    C_G = B_G^T P& \label{CBP_cond_gen} \\
    \tau_r C_G^T C_G \preceq P& \label{CCP_cond_gen}
\end{align}
where $W \triangleq \textrm{blockdiag}\{R,I\}$ and $\tilde{A}_G \triangleq A_G+\tau_s^{-1}I$.
\end{theorem}

\section{Prototype Self-Powered System}\label{sec:proto}

This section describes the physical design, construction, and modeling of the prototype self-powered control system, including identification of parasitic parameters $\{R,\tau_s,\tau_r\}$. 

\subsection{Electromechanical transducer}
\label{sec:transducer}

\begin{figure}
    \centering
   \includegraphics[scale=0.45]{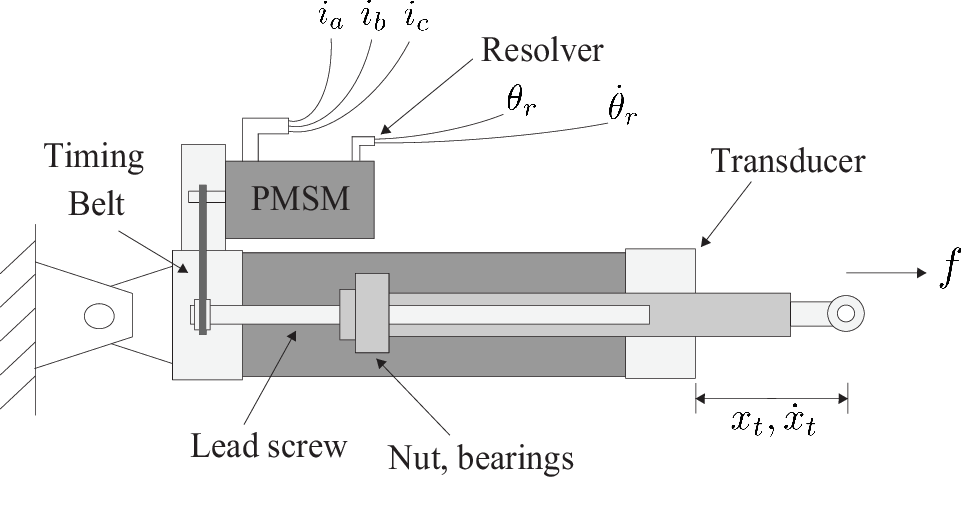}
   \caption{Permanent-magnet synchronous machine (PMSM) transducer with internal components illustrated}
   \label{fig:tran_schematic}
\end{figure}

The transducer consists of a Kollmorgen AKM24C three-phase PMSM, rated at 0.7 kW and 480 V, coupled to a Kollmorgen EC2-series ballscrew actuator via a timing belt. 
An internal resolver in the PMSM provides angular position and velocity measurements. 
A simplified schematic of the transducer is shown in Figure \ref{fig:tran_schematic} and additional data is provided in Table \ref{table:trans_data}.  

\subsubsection{Mechanical dynamic model}

We assume that the linear-to-rotational conversion has negligible backlash, and that the timing belt is rigid, resulting in a proportional relationship between the linear position $x_t$ and the PMSM's mechanical rotation angle $\theta_r$.
The resultant relationship between the respective linear and rotational velocities is ${\dot{x}}_t = \ell \dot{\theta}_r$, where $\ell$ is the effective screw lead length, which includes an amplification factor due to belt drive ratio. 

The total force $f$ produced by the transducer is a result of several interacting physical phenomena.
The rotor has finite rotary inertia $J_t> 0$ and viscous damping $B_t> 0$, which contribute effective rectilinear inertia and damping terms to $f$. 
Additionally, the rectilinear sliding between the ballscrew nut and the guide results in Coulomb friction force $f_c$. 
The electromechanical conversion of energy by the PMSM creates an apparent rectilinear force $f_{em}$ at the ballscrew nut.
The sliding of the bearings between the screw and the nut results in an approximately-static linear-to-rotational conversion efficiency $\eta \in (0,1)$. 
In  \cite{CassidySPIE2011,cassidy2012statistically} it was shown that these effects can be approximately modeled as
\begin{equation}\label{trans_model}
   f = h(p) \left(f_{em} - \frac{J_t}{\ell^2}{\ddot{x}}_t - \frac{B_t}{\ell^2}{\dot{x}}_t\right) - f_c \textrm{sgn}\left({\dot{x}}_t\right)
\end{equation}
where $p$ is the mechanical power delivered to the nut, i.e.,
\begin{equation}
    p = \left(f_{em} - \frac{J_t}{\ell^2}{\ddot{x}}_t - \frac{B_t}{\ell^2}{\dot{x}}_t\right) {\dot{x}}_t , 
\end{equation}
and where $h(\cdot)$ and $\textrm{sgn}(\cdot)$ are 
\begin{align}
    h(p) & \begin{cases}
          = \eta &: \, p > 0 \\
          \in [\eta,1/\eta] &: \, p = 0 \\
          = 1/{\eta} &: \,  p < 0 
     \end{cases}
     \\
     \textrm{sgn}\left({\dot{x}}_t\right) & \begin{cases}
        = 1 &: \, {\dot{x}}_t > 0 \\
        \in [-1,1] &: \, {\dot{x}}_t = 0 \\
        = -1 &: \, {\dot{x}}_t < 0 
     \end{cases}
\end{align}
The mechanical transducer model parameters in equation \eqref{trans_model} were identified using data from a series of characterization experiments, during which the PMSM terminals were left open-circuited.  
A detailed explanation of the identification procedures can be found in chapter 7 of \cite{ligeikis2023self}.

\subsubsection{Electrical dynamic model}

\begin{figure}
\centering
   \includegraphics[scale=0.35]{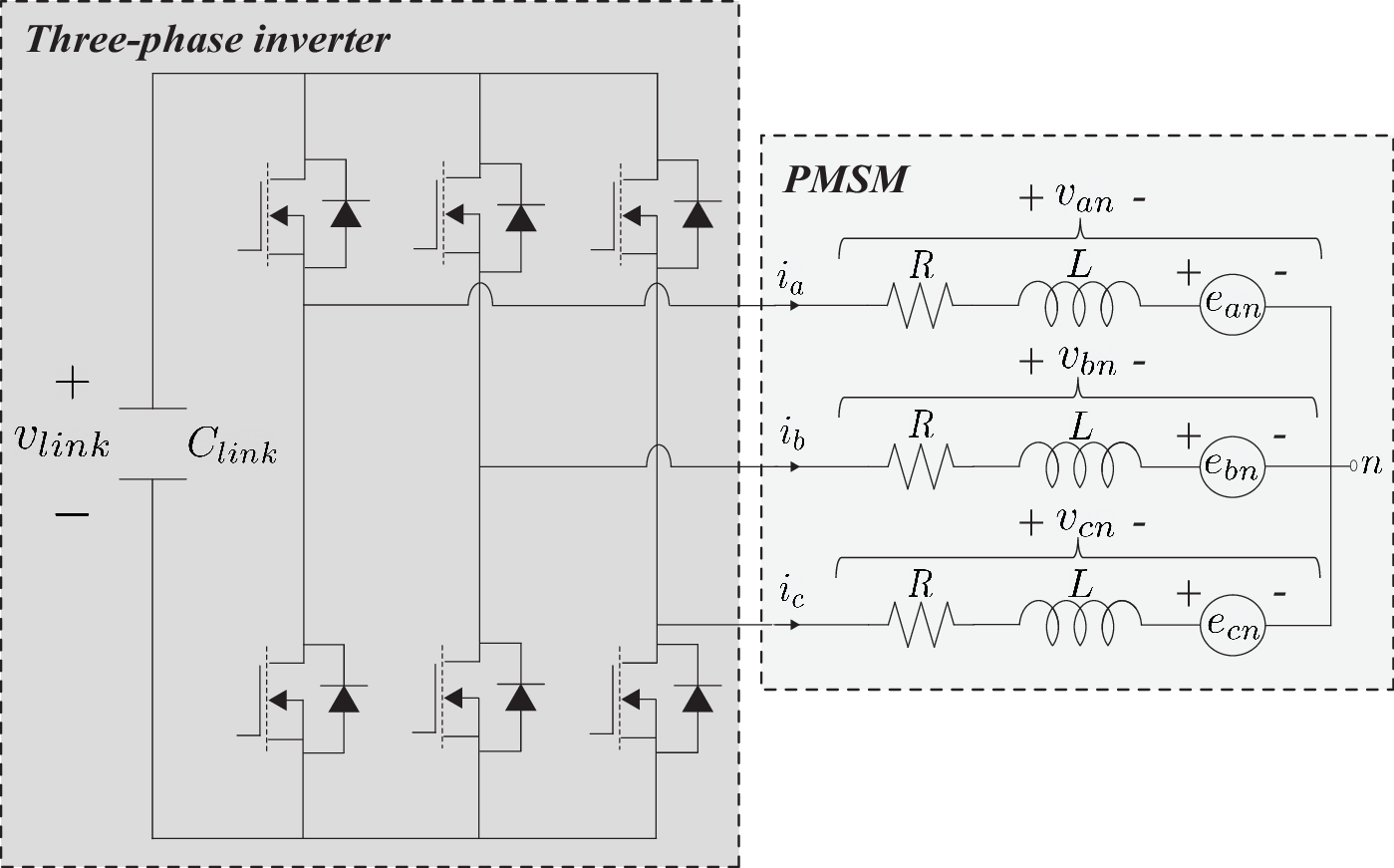}
\caption{Three phase inverter circuit interfaced with permanent-magnet synchronous machine (PMSM)}
\label{fig:inverter}
\end{figure}

The electromechanical force $f_{em}$ depends on the PMSM's three-phase currents, which are described by the differential equation
\begin{equation} \label{stat_eqn}
    \tfrac{d}{dt}i_{abc} = \frac{1}{L} \left( v_{abc} - R_t i_{abc} + e_{abc} \right)
\end{equation}
where $i_{abc}\triangleq [i_a \ i_b \ i_c ]^T$ is the vector of three-phase line-to-neutral currents, $v_{abc}\triangleq[v_{an} \ v_{bn} \ v_{cn}]^T$ is the vector of three-phase line-to-neutral stator voltages, $L$ is the line-to-neutral winding inductance, $R_t$ is the line-to-neutral winding resistance, and $e_{abc}$ is the vector of line-to-neutral back-EMF voltages, found as
\begin{equation}
    e_{abc} \triangleq \begin{bmatrix} e_{an} \\ e_{bn} \\ e_{cn} 
    \end{bmatrix} = \begin{bmatrix} \sin(\theta_{e}) \\ \sin\left(\theta_{e} - \frac{2\pi}{3}\right)  \\ \sin\left(\theta_{e} + \frac{2\pi}{3}\right)  \end{bmatrix} \Lambda_{PM} \dot{\theta}_{e}
\end{equation}
where $\Lambda_{PM}$ is the permanent-magnet flux linkage and $\theta_{e} \triangleq \frac{N_p}{2} \theta_r $ is the electrical rotor angle with $N_p$ being the number of poles of the machine. 
A diagram of the three-phase electrical model of the PMSM is shown in Figure \ref{fig:inverter}.
For the purposes of analysis and control, it is useful to project the three-phrase variables onto a reference frame that rotates with $\theta_{e}$. 
This projection is accomplished using a power-invariant version of the combined  Clarke/Park transformation defined as
\begin{equation} \label{clarke}
    P\left(\theta_{e}\right) \triangleq 
    \setlength{\arraycolsep}{1.5pt}
    \sqrt{\frac{2}{3}}\begin{bmatrix} \cos\left(\theta_{e}\right) & \cos\left(\theta_{e}-\frac{2\pi}{3}\right) & \cos\left(\theta_{e}+\frac{2\pi}{3}\right) \vspace{5pt}\\
    -\sin\left(\theta_{e}\right) & -\sin\left(\theta_{e} - \frac{2\pi}{3}\right) & -\sin\left(\theta_{e} + \frac{2\pi}{3}\right) \vspace{5pt}\\ 
    \sqrt{\frac{1}{2}} & \sqrt{\frac{1}{2}} & \sqrt{\frac{1}{2}}
    \end{bmatrix}
\end{equation}
with the corresponding inverse transformation
\begin{equation}
    P^{-1}\left(\theta_{e}\right) =
    \sqrt{\frac{2}{3}} \begin{bmatrix} \cos\left(\theta_{e}\right) & -\sin(\theta_{e})& \sqrt{\frac{1}{2}}  \vspace{5pt}\\
    \cos\left(\theta_{e} - \frac{2\pi}{3}\right) & -\sin\left(\theta_{e} - \frac{2\pi}{3}\right) & \sqrt{\frac{1}{2}} \vspace{5pt}\\ 
     \cos\left(\theta_{e} + \frac{2\pi}{3}\right) & -\sin\left(\theta_{e} + \frac{2\pi}{3}\right) & \sqrt{\frac{1}{2}}
    \end{bmatrix}.
\end{equation}
Subsequently, we define vectors
\begin{equation}
    i_{dq0}^r \triangleq \begin{bmatrix} i_d^r \\ i_q^r \\ i_0^r
    \end{bmatrix} = P\left(\theta_{e}\right)i_{abc}~,~
    v_{dq0}^r \triangleq \begin{bmatrix} v_d^r \\ v_q^r \\ v_0^r
    \end{bmatrix} = P\left(\theta_{e}\right)v_{abc}
\end{equation}
where the subscripts $dq0$ refer to the direct-axis, quadrature-axis, and zero components, respectively and the superscript $r$ signifies the rotor reference frame. 
It follows that the dynamics of $i_{dq0}^r$ obey the differential equation
\begin{equation}
    \label{r_eqn_init}
    P(\theta_{e})\tfrac{d}{dt}\bigg(P^{-1}(\theta_{e})i_{dq0}^r \bigg)  = \frac{1}{L} \big( v_{dq0}^r - R_t i_{dq0}^r + P(\theta_{e})e_{abc} \big)
\end{equation}
Expanding \eqref{r_eqn_init} and making the substitution $\dot{\theta}_{e} = \frac{N_p}{2\ell}\dot{x}_t$, we obtain the following system of coupled differential equations
\begin{align}
    \tfrac{d}{dt}i_d^r &= \frac{1}{L}\left(v_d^r -R_ti_d^r +\frac{N_p}{2\ell}L i_q^r \dot{x}_t \right) \label{id_eqn} \\
    \tfrac{d}{dt}i_q^r &= \frac{1}{L}\left(v_q^r -R_ti_q^r -\frac{N_p}{2\ell}\left(L i_d^r + \sqrt{\tfrac{3}{2}}\Lambda_{PM}\right) \dot{x}_t  \right) \label{iq_eqn} \\
     \tfrac{d}{dt}i_0^r &= \frac{1}{L}\left(v_0^r -R_ti_0^r  \right) 
\end{align}
The three-phase windings are connected in an ungrounded wye configuration. Applying Kirchoff's current law to the neutral node, we have that $i_0^r = \sqrt{\tfrac{1}{3}}\left(i_a+i_b+i_c\right)  = 0~\forall t$.  
Therefore, it must be the case that $v_0^r = 0 ~\forall t$. 

It can be shown that the electromechanical force produced by the transducer is 
\begin{equation} \label{fe_iq_relationship}
    f_{em} = k_ui_q^r
\end{equation}
where we define 
\begin{equation}
    k_u \triangleq \sqrt{\tfrac{3}{2}}\tfrac{N_p \Lambda_{PM}}{2\ell}.
\end{equation}
Thus by regulating quadrature-axis current $i_q^r$ we can  control the electromechanical force imposed by the transducer on the plant. Since the direct-axis current $i_d^r$ has no effect on the transducer's mechanical dynamics, we control $i_d^r = 0 ~\forall t$, in order to minimize resistive power losses. Following the notation used throughout the previous sections, we let $u \triangleq i_q^r$, and define the colocated voltage to be $v \triangleq k_u \dot{x}_t$.

\subsubsection{Transducer power}
With $i_d^r = 0$, the instantaneous electrical power delivered to the transducer is 
\begin{align}
    P_{e} =& v_{abc}^T i_{abc} 
    = v_{dq0}^{rT} i_{dq0}^r 
    = v_q^r i_q^r  \\
    =& \left( L\tfrac{d}{dt}i_q^r + R_ti_q^r +k_u\dot{x}_t \right)i_q^r \\
    =& \left( L\tfrac{d}{dt}u + R_t u +v \right)u \label{last_p_eqn}
 \end{align}
 Since the PMSM inductance $L$ is small, and the current $u$ varies relatively slowly, it is reasonable to neglect the effect of the first term in equation \eqref{last_p_eqn}. This results in the following approximate expression for $P_e$
 \begin{equation}\label{Pe2}
     P_e \approx R_t u^2 +uv 
 \end{equation}
 where positive $P_{e}$ implies conversion of electrical to mechanical energy. Note that \eqref{Pe2} is just the scalar version of \eqref{Pe1}.

\begin{table}
\caption{Transducer characteristics}
\centering
\small
\label{table:trans_data}
\begin{tabular}{l |c} 
\hline
Parameter & Value\\ 
\hline\hline
Resistance ($R_t$) & 10.6 $\Omega$ \\ 
Inductance ($L$) & 0.0219 H \\
Permanent-magnet flux linkage ($\Lambda_{PM}$) & 0.1603 V-s \\
No. of poles ($N_p$) & 6 \\
Rotational inertia ($J_t$) & 3.54 $\times 10^{-5}$ kg-m$^2$\\ 
Rotational viscous damping ($B_t$) & 3.25 $\times 10^{-4}$ N-m-s\\ 
Coulomb friction ($f_c$) & 35 N\\ 
Lead length ($\ell$) & 1.27 $\times 10^{-3}$ m-rad$^{-1}$\\
Efficiency ($\eta$) & 0.91\\ 
\hline
\end{tabular}
\end{table}

\subsection{Power-Electronic Drives}
\label{sec:power_elecs}

\subsubsection{Three-phase inverter}
\label{subsec:inverter}

\begin{figure}
    \centering
   \includegraphics[scale=0.45]{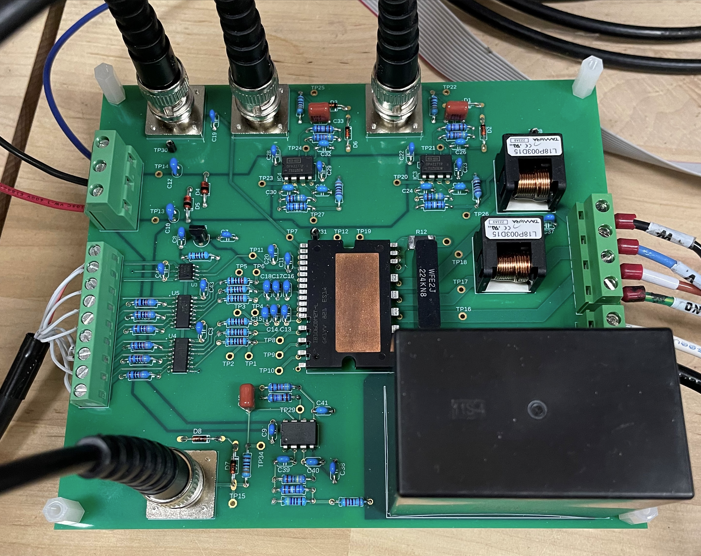}
   \caption{Three-phase inverter circuit}
   \label{fig:inverter_pic}
\end{figure}

A power-electronic drive called a three-phase inverter (see Figure \ref{fig:inverter} for a simplified schematic) is used to regulate the transducer currents. The drive is connected to the DC link voltage, denoted $v_{link}$, and consists of three half-bridge circuits, with the center node of each bridge connected to a line of the PMSM. Each half bridge consists of two transistors (such as MOSFETs or IGBTs) that can effectively be controlled like switches. High-frequency PWM of the transistors allows the drive to generate three-phase AC line-to-line voltages in the PMSM and consequently realize desired phase currents $i_{abc}$. The drive shown in Figure \ref{fig:inverter_pic} was custom designed and built specifically for this research. The circuit is centered around an STMicroelectronics STIB1060DM2T-L intelligent power module (IPM). This IPM consists of a three-phase inverter with MDMesh DM2 MOSFETs, integrated gate drive circuitry, internal bootstrap diodes, fault protection, and a built-in temperature sensor. The MOSFETs have a maximum voltage rating of 600V, a continuous current rating of 12.5A, and a nominal on-state resistance $R_{ON}$ of 0.18$\Omega$. NVE Corporation IL600 Series digital isolators are used to isolate the PWM signals generated by the dSpace DS1103 unit from the inverter circuit. This prevents the formation of ground loops, and the introduction of unwanted noise into the PWM signals. 

A Panasonic EZP-V60117MTS 110$\mu$F metallized polypropylene film capacitor was chosen as the DC link capacitor. This capacitor has a low equivalent series resistance (ESR) to prevent unwanted parasitic losses, and a voltage rating of 600V. A Panasonic ECW-FE2J224QD 0.22$\mu$F metallized polypropylene film capacitor was also placed across the DC link terminals to help prevent voltage spikes from occurring across the MOSFETs during switching transitions. 
The $i_b$ and $i_c$ phase currents are measured using Tamura L18P003D15 Hall-effect sensors, noting that the final phase current can be reconstructed as $i_a=-i_b-i_c$. A voltage divider circuit (with 48:1 attenuation) is used to sense the DC link voltage. To remove high-frequency ripple from the measured measured phase current and DC link voltage signals and prevent aliasing during analog-to-digital conversion, the signals are passed through 4th-order, low-pass Butterworth active filters, which are built using Texas Instruments OPA2277PA operational amplifiers. A cutoff frequency of 2kHz was chosen for these filters. The sensors, signal conditioning circuitry, and IPM gate drivers are powered via a bipolar $\pm$15V supply provided by a Keysight E3631A DC power supply. The total quiescent power required to power the circuit is approximately 0.81W. 

The transducer currents are regulated via proportional-integral (PI) feedback controllers implemented in the rotor reference frame. In this approach, Clarke/Park transformation \eqref{clarke} is applied to the measured three-phase currents $i_{abc}$ to obtain measured rotor reference frame currents $i_q^r$ and $i_d^r$. Error signals are generated by comparing $i_q^r$ and $i_d^r$ to the desired currents $i_q^{r*}$ and $i_d^{r*}$, noting that $i_q^{r*}$ is determined by the self-powered feedback control law, and $i_d^{r*}=0$, as described previously. These error signals are then passed through the PI controllers to produce the necessary $v_q^r$ and $v_d^r$ voltages. An inverse Park transformation is applied to  $v_q^r$ and $v_d^r$ to obtain the two-phase stationary reference frame voltages $v_q$ and $v_r$. Finally, $v_q$ and $v_r$ are fed into a space vector modulation (SVM) algorithm (see e.g., \cite{krishnan2017permanent} for details) which computes the PWM duty ratios for each of the inverter's half-bridges. 

The PWM switching frequency was chosen to be 10kHz. Measurement sampling was synchronized with the PWM switching to prevent the introduction of electromagnetic interference created during switching events. The PI controller gains were tuned to obtain a closed-loop bandwidth of approximately 200Hz, which is roughly two orders of magnitude beyond the dynamics of the plant utilized in our experimental study. We also employed a clamping-type anti-windup scheme to prevent windup of the PI controller integrators.

At a given time $t$, the feasibility of a desired $\{i_q^{r},i_d^{r}\}$ pair depends on $v_{link}$. Since the SVM algorithm is used generate the PWM duty ratios, it can be shown that the maximum magnitude of any three-phase line-to-neutral voltage is $v_{link}/\sqrt{3}$. 
In balanced operation, we have that 
\begin{align}
    v_{an} &= v_{ph} \sin(\theta_{e}+\phi) \\ 
    v_{bn} &= v_{ph} \sin(\theta_{e}-\frac{2\pi}{3}+\phi)\\
    v_{cn} &= v_{ph}\sin(\theta_{e}+\frac{2\pi}{3}+\phi)
\end{align}
where $v_{ph}$ is the voltage amplitude and $\phi$ is an arbitrary constant phase angle. 
Consequently, this implies
\begin{equation} \label{abc_ineq}
    \sqrt{v_{abc}^T v_{abc}}
    = \sqrt{v_{an}^2 + v_{bn}^2 + v_{cn}^2 } = \sqrt{\tfrac{3}{2}} |v_{ph}| 
    \leq \sqrt{\tfrac{3}{2}} \frac{v_{link}}{\sqrt{3}}
\end{equation}
But
\begin{equation} \label{dq0_ineq_1}
    \sqrt{v_{abc}^T v_{abc} } = \sqrt{v_{dq0}^T P^{-T} P^{-1} v_{dq0} } =\sqrt{v_{d}^{r2} + v_{q}^{r2}} 
\end{equation}
Combining \eqref{abc_ineq} and \eqref{dq0_ineq_1} we see that the DC link voltage must satisfy
\begin{equation} \label{dq0_ineq_2}
    \sqrt{2\left(v_{d}^{r2} + v_{q}^{r2}\right)} \leq v_{link} ~~ \forall t
\end{equation}
Next, suppose $\dot{x}_t,~v_q^r,$ and $v_d^r$ vary slowly in time. Since the dynamics of the current control loops are considerably faster than the mechanical dynamics of the system, we have $\tfrac{d}{dt}i_d^r = \tfrac{d}{dt}i_q^r = 0$ in steady-state, which reduces \eqref{id_eqn} and \eqref{iq_eqn} to 
\begin{align} \label{v_d_eqn}
    v_d^r =& Ri_d^r -\frac{N_p}{2\ell}L i_q^r \dot{x}_t
\\
\label{v_q_eqn}
    v_q^r =& Ri_q^r +\frac{N_p}{2\ell}\left(L i_d^r + \sqrt{\tfrac{3}{2}}\Lambda_{PM}\right)\dot{x}_t
\end{align}
Since $i_d^r$ is controlled to be zero, it follows that \eqref{dq0_ineq_2} is equivalent to
\begin{equation} \label{v_link_con}
    \sqrt{2\left(\left(\tfrac{N_p}{2\ell}L i_q^r\dot{x}_t\right)^2 + \left(  Ri_q^r +k_u\dot{x}_t\right)^2\right)}  \leq v_{link} 
\end{equation}
Therefore, to maintain controllability of $\{i_q^{r},i_d^{r}\}$ it is necessary to ensure the DC link voltage always satisfies inequality \eqref{v_link_con}. More details pertaining to DC link voltage regulation are provided in the next section.
 
\subsubsection{Half-bridge DC-DC converter}
\label{subsec:converter}

\begin{figure}
    \centering
   \includegraphics[scale=0.35]{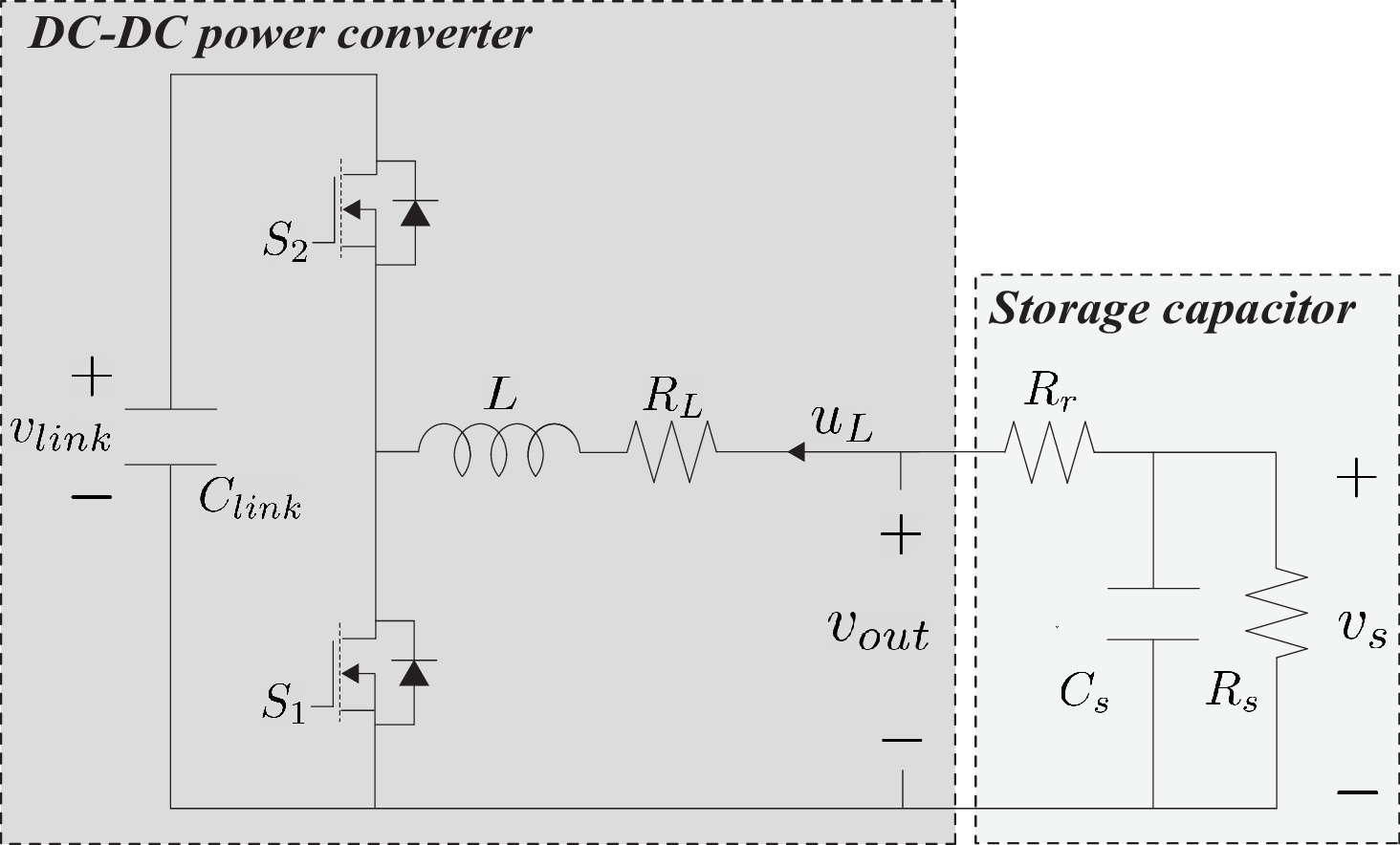}
   \caption{Schematic of half-bridge DC-DC power converter interfaced with storage capacitor}
   \label{fig:converter_schem}
\end{figure}

\begin{figure}%[!htb]
    \centering
   \includegraphics[scale=0.5]{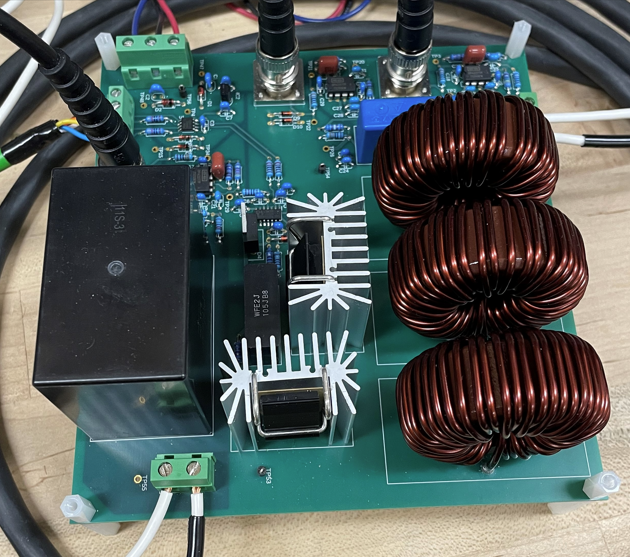}
   \caption{Half-bridge DC-DC power converter circuit}
   \label{fig:dcdc_pic}
\end{figure}

We utilize a half-bridge DC-DC power converter to interface the DC link with the energy storage subsystem. A simplified schematic of the converter is shown in Figure \ref{fig:converter_schem}. It consists of two transistors in a half-bridge configuration, and an inductor $L$ with inherent series resistance $R_L$. The transistors are depicted as MOSFETs in the schematic and labeled $S_1$ and $S_2$. Note that inductor current $u_L = -u_s$. The MOSFETs are controlled synchronously via PWM (i.e., when $S_1$ closes, $S_2$ opens) such that bidirectional power flow is possible between the storage capacitor and DC link capacitor, provided that the DC link voltage is larger than the converter output voltage. (i.e., $v_{link} > v_{out}$). In other words, the drive can act as both a step-down converter and a step-up converter, depending on which way power is flowing. Figure \ref{fig:dcdc_pic} shows the circuit. The half-bridge is comprised of Infineon Technologies IRF200P222 MOSFETs which have a voltage rating of 200V, a continuous current rating of 182A, and a nominal on-state resistance $R_{ON}$ of 5.3m$\Omega$. An Infineon Technologies 2ED21844S06JXUMA1 650V half-bridge gate driver is used facilitate synchronous MOSFET switching. NVE Corporation IL600 Series digital isolators are used to isolate the PWM signals generated by the dSpace DS1103 unit from the converter circuit. A Panasonic EZP-V60117MTC 110$\mu$F metallized polypropylene film capacitor was included to augment the DC link capacitance. A Panasonic ECW-FE2J105JA 1$\mu$F metallized polypropylene film capacitor was added across the DC link terminals to help prevent overvoltages. Three Kemet HHBC24W-2R1A0311V 311$\mu$H inductors with Fe-Si dust toroid cores are used in series to provide a total nominal inductance of 933$\mu$H. The inductors have a combined nominal series resistance of 60.3m$\Omega$ and a DC current rating of 15A. The inductor current is measured using a LEM Model LAH 25-NP Hall-effect sensor, which is powered via a bipolar $\pm$15V supply. A voltage divider circuit (with 23:1 attenuation) is used to sense converter output voltage $v_{out}$. The measured current and voltage signals are again passed through 4th-order, low-pass Butterworth active filters. A cutoff frequency of 2kHz was chosen for these filters. The total quiescent power required to power the circuit is approximately 0.54W. 
 
\begin{figure}
    \centering
   \includegraphics[scale=0.5]{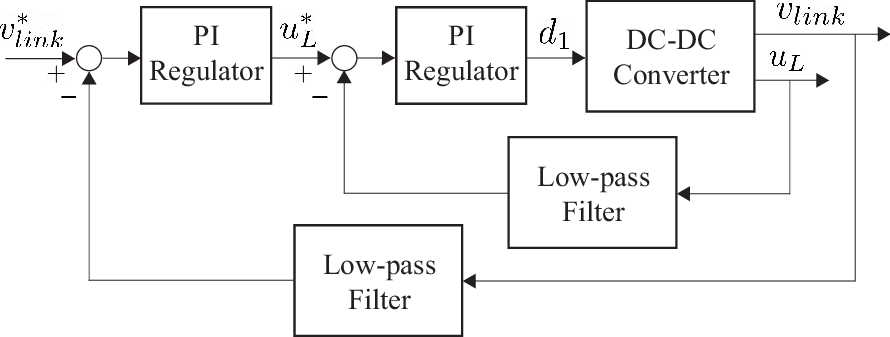}
   \caption{Block diagram of DC-DC power converter control architecture}
   \label{fig:dcdc_block_diagram}
\end{figure}

In our design, the main purpose of the converter is to regulate $v_{link}$ such that transducer current control is feasible (i.e., to satisfy \eqref{v_link_con}) and to prevent $v_{link}$ from growing too large (i.e., beyond the DC link capacitor and MOSFET voltage ratings). To accomplish this, two nested PI regulators are employed as depicted in Figure \ref{fig:dcdc_block_diagram}. The inner PI loop is used to track the reference inductor current $u_L^*$ generated by the outer DC link voltage PI loop. For this approach to work, it is necessary that the inner-loop bandwidth be sufficiently larger than the outer-loop bandwidth.

To obtain an initial design of the PI gains, we used a linearized small-signal model of the converter. Let $d_1$ denote the PWM duty ratio associated with MOSFET $S_1$ and assume the MOSFETs have identical on-state resistance denoted $R_{ON}$. Applying Kirchhoff's voltage law, we obtain the switch-averaged converter dynamics
\begin{equation}
    \left(1-d_1\right)v_{link}+ L \tfrac{d}{dt}u_L + u_L R_L  + u_L R_{ON}= v_{out}
\end{equation}
which are clearly nonlinear due to the product $d_1v_{link}$. Regrouping terms we have
\begin{equation} \label{switch_avg}
    \left(1-d_1\right) v_{link} + u_L \left(R_L+R_{ON}\right) + L \tfrac{d}{dt}U_L = v_{out}
\end{equation}
Next, we assume each signal can be represented as some constant, steady-state value superimposed with a small perturbation, i.e.,
\begin{align}
    d_1 =& D_1 + \tilde{d}_1 \\
    v_{link} =& V_{link} + \tilde{v}_{link} \\
    U_L =& U_L + \tilde{u}_L \\
    v_{out} =& V_{out} + \tilde{v}_{out}
\end{align}
Substituting the previous expressions into \eqref{switch_avg} and regrouping terms we obtain
\begin{multline} \label{perturbed}
    \left[\left(1-D_1\right)V_{link} +U_L \left(R_L+R_{ON}\right) - V_{out} \right] \\ + \bigg[ \left(1-D_1\right)\tilde{v}_{link} - \tilde{d}_1V_{link} + \tilde{u}_L \left(R_L+R_{ON}\right) 
    \\ + L \tfrac{d}{dt} \tilde{u}_L - \tilde{v}_{out}\bigg]  + \left[ \tilde{d}_1\tilde{v}_{link} \right] = 0
\end{multline}
The first bracketed term of \eqref{perturbed} contains only DC signals and thus adds to zero, implying that the steady-state output voltage is given by 
\begin{equation}
    V_{out}  = \left(1-D_1\right)V_{link} +U_L \left(R_L+R_{ON}\right) 
\end{equation}
If $R_L+R_{ON}\approx 0$, then we have $V_{out} = \left(1-D_1\right)V_{link}$. The third bracketed term of \eqref{perturbed} is the product of two perturbation signals, and thus is very small and can be neglected. As such, we are left with the second bracketed term of \eqref{perturbed}, which comprises the small-signal model of the converter
\begin{multline}\label{small_sig_1}
    \left(1-D_1\right)\tilde{v}_{link} - \tilde{d}_1V_{link} + \tilde{u}_L \left(R_L+R_{ON}\right) \\ + L \tfrac{d}{dt} \tilde{u}_L - \tilde{v}_{out} = 0
\end{multline}
Taking the Laplace transform of \eqref{small_sig_1} gives
\begin{equation}
    \hat{u}_L(s) =  \hat{G}_{ud}(s)\hat{d}_1(s) +  \hat{G}_{uv1}(s)\hat{v}_{link}(s) + \hat{G}_{uv2}(s)\hat{v}_{out}(s)
\end{equation}
with
\begin{align}
    \hat{G}_{ud}(s) =& \frac{V_{link}}{sL + R_L + R_{ON}} \\
    \hat{G}_{uv1}(s) =&  \frac{1-D_1}{sL + R_L + R_{ON}} \\
    \hat{G}_{uv2}(s) =& \ -\frac{1}{sL + R_L + R_{ON}}
\end{align}
Note that the gain of  $\hat{G}_{ud}(s)$ is dependent on the steady-state DC link voltage. The system loop gain is given by
\begin{equation}
    \hat{G}_{loop,u}(s) = \hat{G}_{ud}(s)\hat{C_u}(s)\hat{F}(s)
\end{equation}
where $\hat{C_u}(s) = K_{pu} + K_{iu}/s$ is the PI controller transfer function and $\hat{F}(s)$ is the transfer function of the analog 4th-order Butterworth low-pass filter.
Using $\hat{G}_{loop,u}(s)$ we design PI gains $K_{pu}$ and $K_{iu}$ to provide sufficient reference tracking bandwidth, disturbance rejection properties, and gain/phase margins for a range of $V_{link}$ operating points, noting that as $V_{link}$ increases the effective closed-loop bandwidth also increases. Using a similar small-signal analysis, it is straightforward to show that 
\begin{equation}
    \hat{v}_{link}(s) = \hat{G}_{vu1}(s)\hat{u}_{inv}(s) + \hat{G}_{vu2}(s) \hat{u}_L(s)
\end{equation}
where $\hat{u}_{inv}(s)$ denotes the Laplace transform of the perturbed current signal drawn from the DC link capacitor to the three-phase inverter and 
\begin{equation}
     \hat{G}_{vu1}(s) = \frac{1}{s C_s}~,~~~
     \hat{G}_{vu2}(s) = \frac{1-D_1}{s C_s} 
\end{equation}
We note that in the previous expressions it is assumed the ESR of the DC link capacitor is negligible. In addition, the gain of $\hat{G}_{vu2}(s)$ depends on the steady-state duty ratio $D_1$, and if $D_1=1$ then it is not possible to control $v_{link}$. We can again define the loop gain
\begin{equation}
    \hat{G}_{loop,v}(s) = \hat{G}_{vu2}(s) \hat{G}_{rt,u}(s)\hat{C_v}(s)\hat{F}(s)
\end{equation}
where $\hat{C_v}(s) = K_{pv} + K_{iv}/s$ and
\begin{equation}
    \hat{G}_{rt,u}(s) = \frac{\hat{G}_{ud}(s)\hat{C}_u(s)}{1+\hat{G}_{loop,u(s)}}
\end{equation}
is the closed-loop current reference tracking transfer function. Using $\hat{G}_{loop,v}(s)$ we design PI gains $K_{pv}$ and $K_{iv}$ to provide sufficient reference tracking bandwidth, disturbance rejection properties, and gain/phase margins for a range of $\{V_{link},D_1\}$ operating points. In addition, we ensure that there is adequate timescale separation between the inner (current) and outer (voltage) loops. With this preliminary design in hand, the gains $\{K_{pu},K_{iu},K_{pv},K_{iv}\}$ were subsequently tuned based on empirical data to further improve performance.

Lastly, the DC link reference voltage $v_{link}^*$ is determined according to the following equation
\begin{multline} 
    v_{link}^* = \max \Bigg\{ 20,~ v_{out}+\beta, \\\sqrt{2\left(\left(\tfrac{N_pL}{2\ell} i_q^r \dot{x}_t \right)^2 + \left(  Ri_q^r +k_u\dot{x}_t \right)^2\right)} +\beta \Bigg\} \label{v_link_ref}
\end{multline}
where $\beta$ is a constant positive scalar. In our experimental validation, we set $\beta=5$. Thus, if $v_{link} \approx v_{link}^*$, it follows that condition \eqref{v_link_con} is always satisfied and the converter output voltage is always less than the DC link voltage (i.e., ensuring $D_1 < 1$). The constant value of 20V in \eqref{v_link_ref} was chosen to ensure sufficient inductor current tracking bandwidth.

\subsubsection{Analysis of losses in the power-electronic drives and transducer}

The three-phase inverter, DC-DC power converter, and electromechanical transducer exhibit parasitic power loss. These losses are due to a variety of interacting phenomena including conductive (``$I^2 R$") losses, magnetic core losses, and switching losses \cite{erickson2007fundamentals}. The transducer resistance, MOSFET on-state resistance, and the inductor series resistance contribute to the conductive losses. Magnetic losses are comprised of both eddy current losses and hysteresis losses in the inductor cores. Switching losses occur during switching transitions when there is simultaneously significant voltage across and current through a MOSFET. Gate charge losses, deadtime losses, MOSFET body diode reverse recovery losses, and MOSFET output capacitance losses also contribute to the total parasitic loss \cite{erickson2007fundamentals}.

Clearly, the loss model given in \eqref{eq0_1} does not explicitly account for all of these parasitics. However, in this section  we identify a simple, low-order model of the power losses, based on \eqref{eq0_1}, using measured data. 
This model is later augmented with the storage capacitor loss characteristics (see Section \ref{sec:capacitor} below), to obtain a comprehensive approximate loss model. 

\begin{figure}
    \centering
    \includegraphics[scale=0.58,trim={0.5cm 1cm 0cm 1cm},clip]{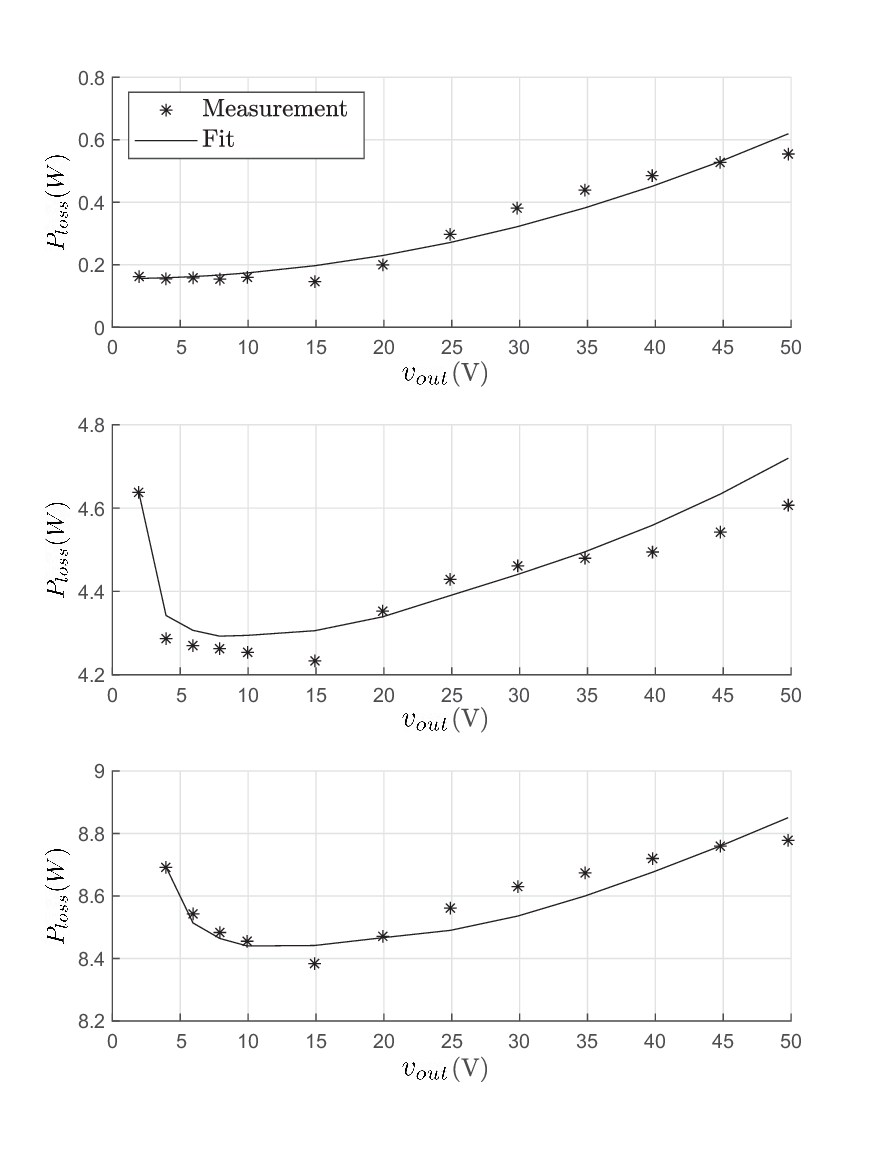}
   \caption{Comparison of measured and modeled parasitic losses for different transducer current conditions: $u=0$ A (top); $u=0.612$ A (middle); and $u=0.866$ A (bottom)}
   \label{fig:drive_loss_model}
\end{figure}

To identify the model, the transducer, inverter, and converter were connected and a DC power supply was used to simulate the storage capacitor. The power supply voltage was varied from 2-50V, and the DC-DC converter was controlled to regulate the DC link voltage according to \eqref{v_link_ref}. During each experiment, the transducer was held stationary and three different transducer quadrature-axis current levels were evaluated: $u=0A,~u=0.612A$, and $u=0.866A$. For each case, the average dissipated power was computed as $P_{loss} = U_LV_{out}$ where $U_L$ denotes the average measured inductor current, and $V_{out}$ is the average measured converter output voltage.  Figure \ref{fig:drive_loss_model} shows the measured power losses. There are a few trends to note. In the top plot (corresponding to $u=0A$), we see that the power losses roughly increase as a quadratic function of $v_{out}$, with some static offset loss. In addition, as $u$ is increased the data are shifted upward due to increased conduction loss. It is important to note that gating losses are not captured in this data, as the gate driver circuitry was powered by the bipolar $\pm$15V supply. However, the gate charge losses were measured separately and found to be less than 0.2W in the worst case.

A least-squares regression was used to fit parameters for the following model
\begin{equation} \label{simple_loss_model}
    P_{loss} = \frac{v_{out}^2}{R_p} + u^2 R_t + u_s^2 R_L + P_0
\end{equation}
where $R_p$ represents an effective ``parallel" resistance and $P_0$ is static power loss term. The identified parameter estimates are $\tilde{R}_p = 5346\Omega$, $\tilde{R}_t=10.96\Omega$, $\tilde{R}_L=0.065\Omega$, and $\tilde{P}_0 = 0.155W$. We note that the identified $\tilde{R}_t$ and $\tilde{R}_L$ match very closely with the nominal combined transducer/inverter MOSFET resistance and combined inductor/converter MOSFET resistance. As shown in Figure \ref{fig:drive_loss_model}, the model captures the general trends of the loss behavior relatively well. 

\subsection{Storage Capacitor}
\label{sec:capacitor}
A Kemet ALS70A104NS100 aluminum electrolytic capacitor, with a nominal capacitance of 0.1 F, was chosen as the energy storage subsystem. This capacitor was partly selected due to its low ESR. The capacitor has a voltage rating of 100V, permitting energy storage up to approximately 500J. A Kemet PYR7511-10 bleeder resistor with a nominal resistance of 10k$\Omega$ and power rating of 13W was secured across the capacitor terminals. This was added for safety purposes to ensure the capacitor is discharged when the system is not in operation.
The resistance was measured using a digital multimeter and determined to be $\tilde{R}_s=9.97$k$\Omega$.

An estimate of leakage time constant $\tau_s$ was determined by charging the capacitor to a constant voltage of 50V using a DC power supply, and then disconnecting the terminals from the supply. The capacitor voltage was then measured as it decayed. Using this data, we estimated the time constant to be $\tilde{\tau}_s=988.4$s. Using the identified $\tilde{R}_s$ and $\tilde{\tau}_s$ parameters, we compute an estimate of $\tilde{C}_s=0.0991$F for the capacitance. 

Several components contribute to the capacitor series resistance $R_r$. First, the capacitor has a nominal ESR of 8m$\Omega$. In addition, the capacitor is connected to the DC-DC converter via 2 feet of AWG16 stranded wire, which contributes another 8m$\Omega$ of resistance. Finally, the terminal block connector on the PCB contributes approximately 5m$\Omega$ of resistance, per its datasheet. Thus, the total series resistance is estimated to be $\tilde{R}_r=0.021\Omega$, implying a power transmission time constant of $\tilde{\tau}_r = 0.0021$s.

\subsection{Summary of Loss Parameters}

In the previous section, we estimated storage capacitor parameters $\tilde{C}_s,\tilde{R}_s,$ and $\tilde{R}_r$. These estimates did not consider the effect of additional parasitic losses introduced by the power-electronic drives. Accordingly, in this section we modify these estimates using drive loss model \eqref{simple_loss_model} to obtain the final loss parameters for our prototype self-powered system. Clearly, $\tilde{R}_r$ and $\tilde{R}_L$ are in series. As such, we can simply add them together to obtain $R_r = 0.086\Omega$ and $\tau_r =0.0085$s. If we assume that $v_{out} \approx v_s$, which is not unreasonable considering that $R_r$ is relatively small, we can combine parallel resistances $\tilde{R}_p$ and $\tilde{R}_s$, giving $R_s = 3480\Omega$.
This implies $\tau_s = 344.9$s. Finally, we assume $R=\tilde{R}_t=10.96\Omega$. In summary, the dissipative losses in our system are approximately 
\begin{equation} \label{final_loss_equation}
    P_d \approx u^2R +  u_s^2 R_r + \tfrac{1}{R_s}v_s^2 + P_0
\end{equation}
We note that \eqref{final_loss_equation} is a slightly modified version of \eqref{eq0_1}. A list of the loss parameter values is provided in Table \ref{table:loss_table}.

\section{Example Plant}
\label{sec:rths_example_system}

\begin{figure}
    \centering
   \includegraphics[scale=1.3,trim={0cm 0 0cm 0},clip]{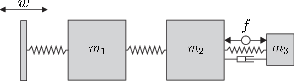}
   \caption{Illustration of 3-DOF vibratory system considered in HiL testing and numerical simulations}
   \label{fig:3dof}
\end{figure}

To demonstrate the capabilities of the control hardware prototype described in the previous section, HiL experiments were conducted in which the prototype is used to control a virtual vibrating structure as shown in Figure \ref{fig:3dof}. The structure is a two-mass LTI system with a tuned vibration absorber (TVA) attached to $m_2$. Masses $m_1$ and $m_2$ are each 75000kg, and without the TVA, the structure natural frequencies are approximately 0.62Hz and 1.63Hz, with 1$\%$ modal damping for each mode. The TVA mass $m_3$ is 3000kg, which is 2$\%$ of $m_1+m_2$. The system is excited by stochastic base acceleration disturbance $\ddot{x}_b$. The natural frequency of the TVA was chosen to be 1Hz, which corresponds to the center frequency of the assumed disturbance model (which is described subsequently), with a damping ratio of 0.1$\%$. 

The dynamics of this system are governed by the following matrix differential equation
\begin{equation} \label{structure_dynamics}
    M_b \ddot{q} + C_b \dot{q} + K_b q = \Gamma_f f - M_b \bar{1}\ddot{x}_b
\end{equation}
where $q=[x_1 \ x_2 \ x_3 ]^T$ is a vector of mass displacements relative to the base, $\ddot{x}_b$ is the base acceleration, $f$ is the transducer force, $\Gamma_f \triangleq [0 \ -1 \ 1 ]^T$, $\bar{1} \triangleq [ 1 \ 1 \ 1 ]^T$, and 
\begin{align}
    M_b =& \begin{bmatrix}
        75000 & 0 & 0 \\
        0 & 75000 & 0 \\
        0 & 0 & 3000
    \end{bmatrix}\textrm{kg} \\
    C_b =& \begin{bmatrix}
        12728 &	-4243 &	0 \\
        -4243 &	8523 &	-37.95 \\
        0 &	-37.95 &	37.95
    \end{bmatrix}\textrm{Ns/m} \\
    K_b =& \begin{bmatrix}
        6000000 &   -3000000   &        0 \\
        -3000000  &   3120000  &   -120000 \\
           0   &  -120000   &   120000
    \end{bmatrix}\textrm{N/m}
\end{align}
Recalling that $f$ is as in \eqref{trans_model}, we have \eqref{structure_dynamics} as
\begin{multline} 
\label{structure_dynamics_2}
    M_b \ddot{q} + C_b \dot{q} + K_b q = -M_b \bar{1}\ddot{x}_b 
    - \Gamma_f f_c \textrm{sgn}\left(\Gamma_f\dot{q}\right)  \\    
    + \Gamma_f h(p)  \left(f_{em} - \tfrac{J_t}{\ell^2}\Gamma_f^T\ddot{q} - \tfrac{B_t}{\ell^2}\Gamma_f^T\dot{q}\right) 
\end{multline}
Then using \eqref{fe_iq_relationship} and rearranging terms, we obtain
\begin{multline} \label{structure_dynamics_3}
    \left(M_b + h(p)\tfrac{J_t}{\ell^2}\Gamma_f \Gamma_f^T\right) \ddot{q}  +\left(C_b + h(p)\tfrac{B_t}{\ell^2}\Gamma_f \Gamma_f^T \right) \dot{q} + K_b q \\  = \Gamma_f h(p)  k_u u  
    - \Gamma_f f_c \textrm{sgn}\left(\Gamma_f\dot{q}\right) - M_b \bar{1}\ddot{x}_b 
\end{multline}
which is nonlinear due to the efficiency function $h(p)$ and the Coulomb friction force.

\begin{table} 
\caption{Prototype self-powered system loss parameters}
\centering
\small
\label{table:loss_table}
\begin{tabular}{l c} 
Parameter & Value\\ 
\hline\hline
Leakage time constant ($\tau_s$) & 344.9 s \\
Transmission time constant ($\tau_r$) & 0.0085 s \\
Resistance ($R$) & 10.96 $\Omega$ \\ 
Capacitance ($C_s$) & 0.0991 F \\ 
Leakage resistance ($R_s$) & 3480 $\Omega$ \\
Series resistance ($R_r$) & 0.086 $\Omega$ \\ 
Static power loss ($P_0$) & 0.155 W \\
\hline
\end{tabular}
\end{table}

To make controller synthesis analytically tractable, we assume that the transducer is backdriven for the majority of the dynamic response, resulting in $p<0$ (i.e., negative mechanical power delivered to the ballscrew nut) for most $t$.
As such, for the purposes of control design, the function $h(p)$ can be approximated by $1/\eta$ for all $t$, which simplifies \eqref{structure_dynamics_3} to
\begin{equation} \label{structure_dynamics_4}
    \tilde{M} \ddot{q} + \tilde{C} \dot{q} + K_b q =  \Gamma_f  k_u u - \Gamma_f f_c \textrm{sgn}\left(\Gamma_f\dot{q}\right) - M_b \bar{1}\ddot{x}_b 
\end{equation}
where the linearized mass and damping matrices are 
\begin{equation}
    \tilde{M} \triangleq M_b + \tfrac{J_t}{\eta\ell^2}\Gamma_f \Gamma_f^T,~~~ \tilde{C} \triangleq C_b + \tfrac{B_t}{\eta\ell^2}\Gamma_f \Gamma_f^T 
\end{equation}
In state-space form, \eqref{structure_dynamics_4} becomes
\begin{equation} \label{structure_ss}
    \tfrac{d}{dt}{x}_p = A_p x_p + B_{pu} u + F_p\textrm{sgn}(\Gamma_{pv} x_p) + B_{pw} w
\end{equation}
where $x_p = \begin{bmatrix} q^T & \dot{q}^T\end{bmatrix}^T$ and $w \triangleq \ddot{x}_b$ and
\begin{equation}
    A_p \triangleq \begin{bmatrix}
        \bar{0} & I \\
        -\tilde{M}^{-1}K_b & -\tilde{M}^{-1}\tilde{C}
    \end{bmatrix}, ~~~
    B_{pu} \triangleq \begin{bmatrix} \bar{0}\\ \tilde{M}^{-1}\Gamma_f k_u \end{bmatrix}
\end{equation}
\begin{equation}
    B_{pw} \triangleq \begin{bmatrix}
        \bar{0} \\ -\tilde{M}^{-1}M_b \bar{1}
    \end{bmatrix},~~~ F_p \triangleq \begin{bmatrix}
        \bar{0} \\ -\tilde{M}^{-1}\Gamma_f f_c
    \end{bmatrix}
\end{equation}
\begin{equation}
    \Gamma_{pv} \triangleq \begin{bmatrix}
        \bar{0} & \Gamma_f^T
    \end{bmatrix}
\end{equation}
where $\bar{0}$ denotes an appropriately sized matrix of zeros. The colocated voltage output is $v = C_{pv}x_p$ where
\begin{equation}
    C_{pv} \triangleq k_u\Gamma_{pv}
\end{equation}
We note that \eqref{structure_ss} is still nonlinear due to the Coulomb friction force.
in Section \ref{subsec:stoch_lin} we employ stochastic linearization \cite{roberts2003random} to accommodate this nonlinearity in control design.

\begin{figure}%[!htb]
    \centering
   \includegraphics[scale=0.55,trim={0.3cm 16.5cm 1cm 1.2cm},clip]{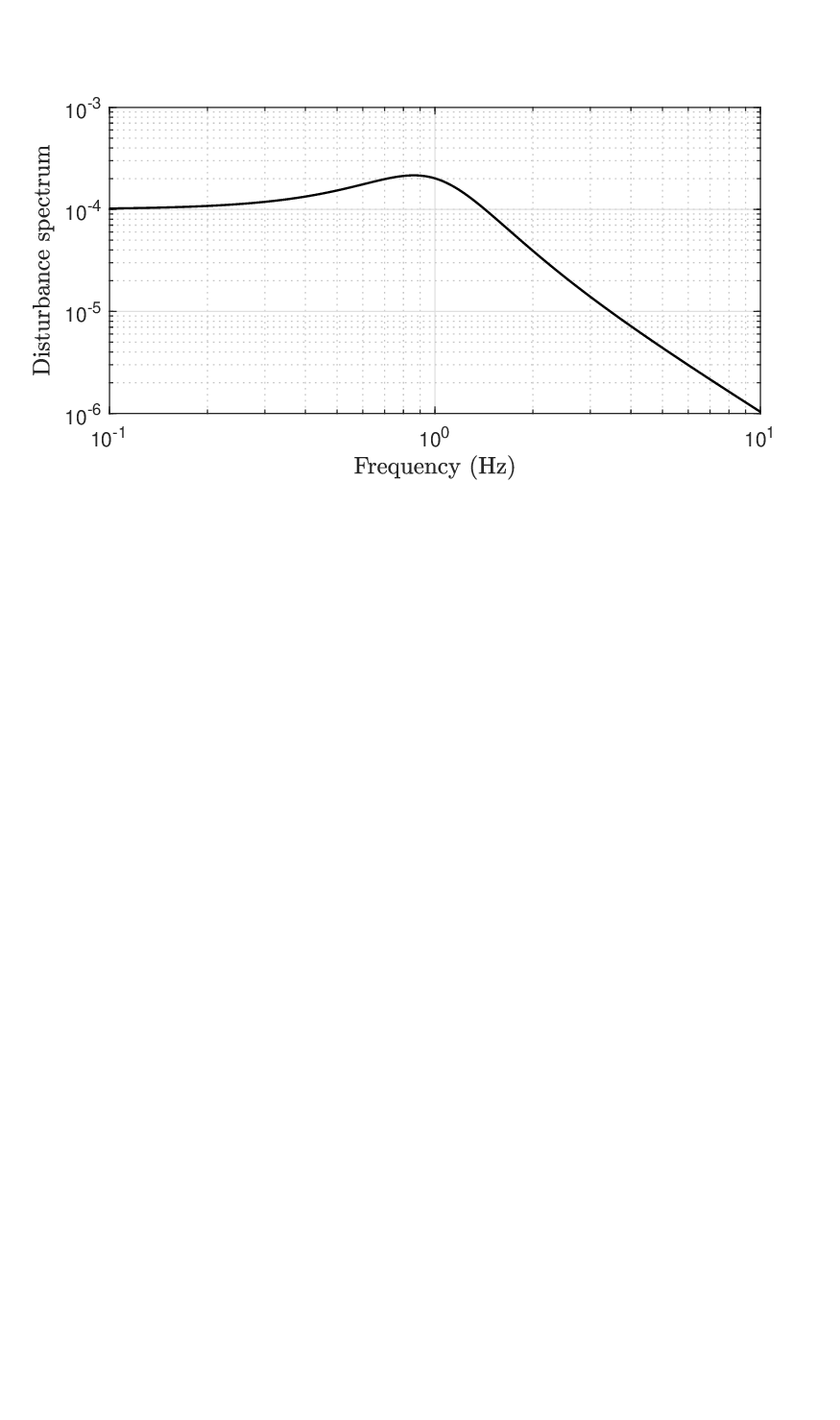}
   \caption{Power spectral density of base acceleration disturbance}
   \label{fig:psd}
\end{figure}

We model the base acceleration as the output of a second-order filter with state-space representation
\begin{equation} 
\mapp{W}: \left\{ \begin{array}{rl}
\tfrac{d}{dt}x_w =& A_w x_w + B_w n \\
                     w =& C_w x_w
\end{array} \right.
\label{KT_filter}
\end{equation} 
where $n$ is a zero-mean, stationary white noise process with unit spectral intensity, $x_w$ is the filter state vector, and
\begin{equation} 
\ A_w \triangleq \begin{bmatrix} 0 & 1 \\ -\omega_w^2 & -2\zeta_w\omega_w \end{bmatrix},~~~ B_w \triangleq \begin{bmatrix} 0 \\ \sigma_w \end{bmatrix}
\end{equation}
\begin{equation} 
 C_w \triangleq \begin{bmatrix} \omega_w^2 & 2\zeta_w\omega_w \end{bmatrix}
\end{equation}
We assume a natural frequency of $\omega_w=2\pi$ rad/s, a damping ratio of $\zeta_w=0.5$, and an intensity of $\sigma_w=0.01$m/s$^2$. As shown in Figure \ref{fig:psd}, the disturbance spectrum has a low quality factor and is centered at the natural frequency of the TVA.

We augment plant model \eqref{structure_ss} and disturbance model \eqref{KT_filter} to obtain the nonlinear, state-space model
\begin{equation} 
\mapp{S}: \left\{ \begin{array}{rl}
\tfrac{d}{dt}x =&\hspace{-5pt} A x + B_u u + F\textrm{sgn}(\Gamma_v x)+ B_n n \\
v =& \hspace{-5pt} C_v x \\
z =& \hspace{-5pt} C_z x + D_{zu}u
\end{array} \right.
\label{x_deq}
\end{equation} 
where the composite state vector $x \triangleq [ x_p^T \ x_w^T ]^T$, $z$ is a vector of performance outputs, and the various parameter matrices are appropriately defined.

\section{Controller Design}
\label{sec:rths_example_control_design}

In this section, we describe the design of self-powered feedback controllers which were evaluated via HiL testing. The controller performance is benchmarked by comparing to optimal static damping. To account for uncertainty in the estimated self-powered system loss parameters given in Table \ref{table:loss_table}, we conservatively use $\bar{R}=11\Omega$, $\bar{\tau}_s =  275$s, and $\bar{\tau}_r=0.01$s for the purposes of controller synthesis. Also, we note that the sufficient feasibility conditions derived in \cite{ligeikis2023feasibility} and the design methodologies we employ herein do not account for the static power loss term $P_0$ in \eqref{final_loss_equation}. Indeed, the design methodology outlined in this section neglects the effect of static parasitic losses. This simplification is justified by the small value identified for $P_0$ in the previous section, and also by the fact that upon experimental implementation and testing, it was found that controllers designed under this simplifying assumption remained feasible and performed as expected.  

\subsection{Performance objective}

We seek to minimize the mean-square performance objective $J$, defined as
\begin{equation}
    J \triangleq \Ex \left\{ z^T z \right\} 
    =\Ex \left\{ x^T Q x + u^T M u + 2x^T N u \right\}
\end{equation}
where $\Ex\{\cdot\}$ denotes expectation in stationarity and $Q \triangleq C_z^T C_z$, $M \triangleq D_{zu}^T D_{zu}$, and $N \triangleq C_z^T D_{zu}$.

For performance outputs $z$ we select the absolute accelerations of masses $m_1$ and $m_2$ and the transducer current $u$ with the following weightings
\begin{equation}
    z_1 = \ddot{x}^{abs}_1,~~~~~~ z_2 = \ddot{x}^{abs}_2,~~~~~~ z_3 = 0.0286 u.
\end{equation}

\subsection{Stochastic linearization of transducer dynamics}
\label{subsec:stoch_lin}

To facilitate controller design, we linearize $\mapp{S}$, as defined in \eqref{x_deq} above. We presume the following LTI, colocated feedback mapping $\mapp{Y}:v\mapsto -u$ is imposed
\begin{equation} \label{Y_controller}
    \mapp{Y}: \left\{ \begin{array}{rl} \tfrac{d}{dt}x_Y &= A_Y x_Y + B_Y v \\
    -u &= C_Y x_Y + D_Y v
    \end{array} \right.
\end{equation}
We assume that, in closed-loop, the augmented state vector
\begin{equation}
    \nu = \begin{bmatrix} x^T & x_Y^T \end{bmatrix}^T
\end{equation}
has a probability distribution $\phi(\nu)$ that can be approximated as Gaussian with zero mean and stationary covariance matrix $\Sigma = \Ex \{\nu \nu^T \}$, i.e., 
\begin{equation}
    \phi(\nu) \approx \frac{1}{\sqrt{(2\pi)^n \det \Sigma}}\exp\left\{-\frac{1}{2}\nu^T \Sigma \nu\right\}
\end{equation}
and then find the value of $\Sigma$ that enforces the weak stationarity condition
\begin{equation}
    \tfrac{d}{dt} \Ex\left\{ \nu\nu^T \right\} = 0
\end{equation}
It can be shown (see \cite{cassidy2012statistically} for details) that this results in the solution to the nonlinear, Lyapunov-like equation
\begin{equation} \label{lyap_like}
    A_{cl}(\Sigma)\Sigma + \Sigma A_{cl}^T(\Sigma) + B_{ncl} B_{ncl}^T = 0
\end{equation}
where 
\begin{equation} \label{A_cl_eqn}
    A_{cl}(\Sigma) \triangleq \begin{bmatrix} A_{eq}(\Sigma)-B_u D_Y C_v & -B_u C_Y \\ B_Y C_v & A_Y \end{bmatrix}, ~~ 
        B_{ncl} = \begin{bmatrix} B_n \\ 0 \end{bmatrix}
\end{equation}
and where
\begin{equation} \label{V_eq}
    A_{eq}(\Sigma) = A + \sqrt{\frac{2}{\pi}}\frac{F\Gamma_v}{\sqrt{\Gamma_v \Sigma \Gamma_v^T}}
\end{equation}
We then have the stochastically-linearized model 
\begin{equation}
\mapp{S}_{eq}: \left\{ \begin{array}{rl}
    \frac{d}{dt}x &= A_{eq}(\Sigma) x + B_u u  + B_n n \\
    v &= C_v x \\
    z &= C_z x 
\end{array} \right.
\label{lin_state_space}
\end{equation}
We emphasize that $\mapp{S}_{eq}$ implicitly depends on the imposed feedback law $\mapp{Y}$ given the relationship between $A_{eq}$ and $\Sigma$. Thus, an iterative technique is required to design $\mapp{Y}$.

\subsection{Optimized static damping}

We first design a controller of the form
\begin{equation} \label{visc_damp}
    u = -c_d v
\end{equation}
where $c_d \geq 0$ is a constant scalar parameter. Feedback law \eqref{visc_damp} is analogous to imposing a synthetic resistance across the terminals of the transducer.
It may also be viewed as approximately equivalent to the imposition of synthetic mechanical viscous damping, since $u$ and $v$ are proportional to the transducer force and colocated velocity, respectively. We seek to optimize the damping coefficient $c_d$ such that $J$ is minimized. 
It is straightforward to show that the closed-loop performance is 
\begin{equation} \label{damp_perf}
    J = \trace \{C_{zcl}\Sigma C_{zcl}^T \}
\end{equation}
where again $\Sigma$ is the solution to equation \eqref{lyap_like} with $A_{cl}(\Sigma) = \left(A_{eq}(\Sigma)-B_u c_d C_v \right)$ and $C_{zcl} = C_z-D_{zu} c_d C_v$.

The optimization problem is defined as
\begin{equation}\label{opt_damping}
     \begin{array}{lll}
    \text{Given:} && \mapp{S}_{eq} \\
    \text{Minimize:} && \eqref{damp_perf} \\
    \text{Over:} &&  c_d \in [0 , \bar{R}^{-1} ]
    \end{array} 
\end{equation}
which can be solved easily. We note the domain constraint in \eqref{opt_damping} is equivalent to sufficient condition $\eqref{Z_condition}$. As previously stated, since $\mapp{S}_{eq}$ is a function of $\Sigma$, which in turn depends on $c_d$, it is necessary to iteratively solve \eqref{opt_damping}. As such, we propose the following simple procedure to optimize $c_d$ for the stochastically-linearized system model:
\begin{itemize}
  \item[] \textbf{Step 0.} ~~Set $A_{eq}=A$, and solve \eqref{opt_damping} to get initial $c_d$.
  \item[] \textbf{Step 1.} Assemble $A_{cl}$ as in \eqref{A_cl_eqn} and compute $\Sigma$ by solving equation \eqref{lyap_like}.
  \item[] \textbf{Step 2.} Compute $A_{eq}(\Sigma)$ via \eqref{V_eq}.
  \item[] \textbf{Step 3.} Re-solve \eqref{opt_damping} for the updated $\mapp{S}_{eq}$ to obtain new $c_d$. Return to Step 1.
\end{itemize}
Steps 1-3 are repeated until performance measure $J$ has converged. Following this procedure converges to $c_d^*=0.06722 \Omega^{-1}$ with $J(c_d^*) = 0.0243$.

\subsection{LTI SPSA}

\begin{figure}
    \centering
   \includegraphics[scale=0.55]{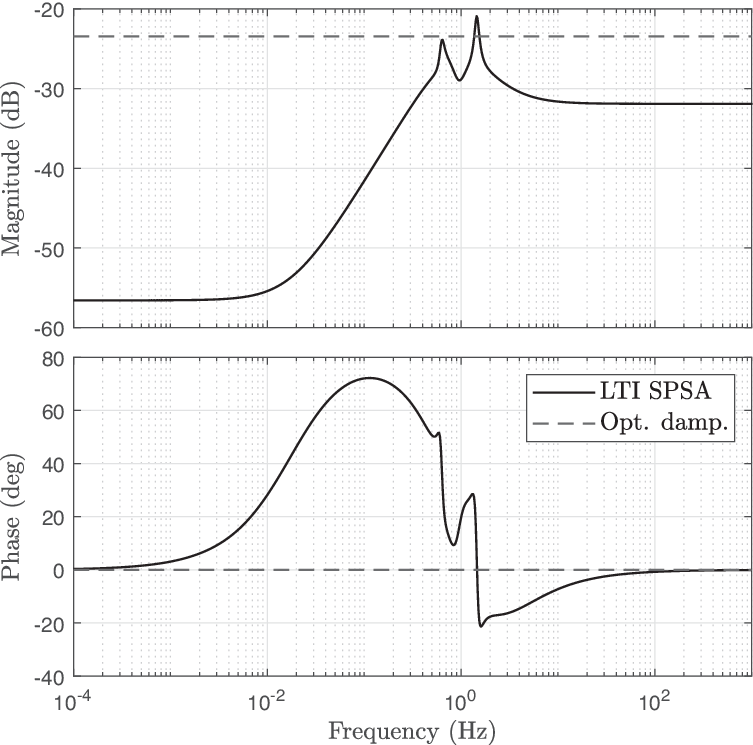}  %,trim={0.1cm 9.1cm 1cm 1.2cm},clip]
   \caption{Bode plot of LTI SPSA controller and optimal static damping}
   \label{fig:spsa_bode}
\end{figure}

Next, we design an LTI SPSA controller using the methodology proposed by the authors in \cite{ligeikis2022lqg}. The method consists of two phases. First, a \emph{passive} LTI feedback law $\mapp{C}: v\mapsto -u$ is designed such that $J$ is minimized. This involves solving a nonconvex, but tractable, optimization. Then, a self-powered feasible controller is obtained by projecting $\mapp{C}$ onto the set $\mathbb{Y}_{1}(R,\tau_s,\tau_r)$. To be more specific, we find the LTI SPSA $\mapp{Y}$ that best approximates $\mapp{C}$ using $\|(\hat{Y}(s)-\hat{C}(s))\hat{P}_{uv}(s)\|_2$ as the error metric, where $\hat{P}_{uv}(s)$ is the transfer function for the open-loop plant mapping $u \mapsto v$. This projection is conservatively formulated as a convex optimization. Due to space constraints, we do not fully describe the method here, but instead direct the reader to \cite{ligeikis2022lqg} for more details. An iterative procedure, analogous to the one described in the previous section, was used to account for the effect of Coulomb friction. Figure \ref{fig:spsa_bode} shows a Bode plot of the resulting SPSA, along with the optimal static damping gain. The analytical closed-loop performance with $\mapp{Y}$ imposed is $J(\mapp{Y})= 0.0200$, which constitutes about an 18$\%$ improvement in performance relative to optimal static damping.

\subsection{Nonlinear performance-guaranteed controller}

Finally, we design a nonlinear self-powered controller using a methodology known as Performance-Guaranteed Control (PGC) \cite{scruggs2007non,scruggs2007non2,scruggs2010multi,cassidy2013nonlinear,ligeikis2021nonlinear,ligeikis2022lqg}. Although sub-optimal, the PGC technique has the advantage of guaranteeing to improve upon the performance of the LTI SPSA designed in the previous section, while ensuring self-powered feasibility. We first recall a theorem from \cite{ligeikis2022lqg} that shows how a time-invariant $Z_0$ matrix and LTI system $\mapp{G}$ that is equivalent with a given LTI SPSA $\mapp{Y}$ may be derived.

\begin{theorem}[\cite{ligeikis2022lqg}] \label{equiv_theorem}
    Let $\mapp{Y}$ with realization \eqref{Y_controller} and matrices $P$ and $X$ satisfy the conditions of Corollary \ref{lti_main_theorem}. Then an equivalent representation of $\mapp{Y}$, as in Figure \ref{fig:LFT}, in which $\mapp{G}$ is characterized by \eqref{G_equation}, is given by
    \begin{align*}
        A_G =& A_Y + P^{-1}X,~~~~~~~~~
        B_G = \sqrt{\tau_r^{-1}}P^{-1}U\Psi^{1/2}V \\
        C_G =& \sqrt{\tau_r^{-1}}V^T\Psi^{1/2}U^T,~~~
        Z_{11} = D_Y \\
        Z_{12} =& \sqrt{\tau_r} C_Y U \Psi^{-1/2} V,~~~~
        Z_{21} = -\sqrt{\tau_r}V^T \Psi^{-1/2} U^T P B_Y \\
        Z_{22} =& \tau_r V^T \Psi^{-1/2}U^T X U \Psi^{-1/2} V 
    \end{align*}
    where $U\Psi U^T$ is the singular value decomposition of positive definite matrix $P$ and $V$ is any unitary matrix (i.e., $VV^T = I)$.
\end{theorem}

Next, we define augmented state vector  $\bar{x} \triangleq [ x^T \ x_G^T ]^T$, output vector $\bar{v}^T \triangleq [ v^T \ r^T ]^T$, input vector  $\bar{u}^T \triangleq [ u^T \ q^T ]^T$, and state space parameter matrices
\begin{align}
\bar{A} =& \begin{bmatrix} A_{eq} & 0 \\ 0 & A_G \end{bmatrix}, &
\bar{B}_u=& \begin{bmatrix} B_u & 0 \\ 0 & B_G \end{bmatrix}, &
\bar{B}_n=& \begin{bmatrix} B_n \\ 0 \end{bmatrix} \\
\bar{C}_v=& \begin{bmatrix} C_v & 0 \\ 0 & C_G \end{bmatrix}, &
\bar{C}_z=& \begin{bmatrix} C_z & 0 \end{bmatrix}, &
\bar{D}_{zu} =& \begin{bmatrix} D_{zu} & 0 \end{bmatrix} 
\end{align}
It follows that with the LTI SPSA $\mapp{Y}$ feedback law imposed, the closed-loop system has state-space realization
\begin{align}
\tfrac{d}{dt}\bar{x} =& \left( \bar{A} - \bar{B}_u Z_0 \bar{C}_v \right) \bar{x}  + \bar{B}_n n \\
z =& \left( \bar{C}_z - \bar{D}_{zu}Z_0 \bar{C}_v \right) \bar{x}
\end{align}
where time-invariant matrices $Z_0,A_G,B_G,C_G$ are obtained by applying the results of Theorem \ref{equiv_theorem} to $\mapp{Y}$. It can be shown that the closed-loop performance of the LTI SPSA $\mapp{Y}$ is 
\begin{equation} \label{Y_perf}
    J(\mapp{Y}) = \trace{\{\bar{B}_n^T P_Y \bar{B}_n\}}
\end{equation}
where matrix $P_Y$ is the solution to the Lyapunov equation
\begin{multline}\label{P_lyap}
    \left( \bar{A} - \bar{B}_u Z_0 \bar{C}_v \right)^T P_Y + P_Y\left( \bar{A} - \bar{B}_u Z_0 \bar{C}_v \right) + \bar{Q} \\ + \bar{C}_v^T Z_0^T \bar{M} Z_0 \bar{C}_v - \bar{C}_v^T Z_0^T \bar{N}^T - \bar{N} Z_0 \bar{C}_v =0
\end{multline}
with $\bar{Q},\bar{M},\bar{N}$ defined as
\begin{equation}
    \bar{Q} \triangleq \begin{bmatrix}
        Q & 0 \\ 0 & 0
    \end{bmatrix}, ~~~\bar{M} \triangleq \begin{bmatrix}
        M & 0 \\ 0 & 0
    \end{bmatrix},~~~ \bar{N} \triangleq \begin{bmatrix}
        N & 0 \\ 0 & 0
    \end{bmatrix}.
\end{equation}

We now formulate a nonlinear, full-state feedback control law $\mapp{K}: \bar{x}\mapsto\bar{u}$, in which a convex optimization problem is solved in real-time. The below controller was originally proposed in \cite{ligeikis2022lqg}, and was proven to result in performance $J(\mapp{K}) \leq J(\mapp{Y})$ while ensuring self-powered feasibility.

\begin{equation}\label{of_controller2}
    \ \bar{u}(t) = \underset{\tilde{\bar{u}}}{\textrm{sol}} \left\{ \begin{array}{ll}
    \text{Given}:& \bar{x}(t), P_Y \\
    \text{Minimize}:& \tilde{J} \\
    \text{Over}:& \tilde{J}, \tilde{\bar{u}} \\
    \text{Subj. to} :& \tilde{\bar{u}}^T(t) W \tilde{\bar{u}}(t) + \bar{x}^T(t) \bar{C}_v^T \tilde{\bar{u}}(t) \leq 0    \end{array} \right.
\end{equation}
where $\tilde{J} \triangleq \tfrac{1}{2}\tilde{\bar{u}}^T \bar{M} \tilde{\bar{u}} +\bar{x}^T(t)\left(P_Y \bar{B}_u + \bar{N}\right)\tilde{\bar{u}}$.

Note that $\bar{M} \succeq 0$ and $W \succ 0$, implying that \eqref{of_controller2} is convex. Next, we show that the optimization in \eqref{of_controller2} can actually be reduced to a root solving problem via application of the Karush-Kuhn-Tucker (KKT) conditions. Without loss of generality, we assume that $\bar{x}$ is not in the nullspace of matrix $\bar{C}_v$ (i.e., $\bar{C}_v\bar{x} \neq 0)$. If $\bar{C}_v\bar{x}=0$, then clearly the only feasible solution to \eqref{of_controller2} is $\tilde{\bar{u}}=0$.
We formulate the Lagrangian function
\begin{multline}
    \label{lagrangian}
    \mathcal{L}(\tilde{\bar{u}},\mu) = \bar{x}^T\left(P_Y \bar{B}_u +\bar{N}\right)\tilde{\bar{u}} + \frac{1}{2} \tilde{\bar{u}}^T \bar{M} \tilde{\bar{u}} \\ + \mu \left( \tilde{\bar{u}}^T W \tilde{\bar{u}} + \bar{x}^T \bar{C}_v^T \tilde{\bar{u}}\right)
\end{multline} 
where $\mu$ is the scalar dual variable associated with inequality constraint \eqref{u_cond}. We compute the gradient of \eqref{lagrangian} with respect to $\tilde{\bar{u}}$ to obtain
\begin{equation}
    \nabla_{\tilde{\bar{u}}}\mathcal{L}= \bar{x}^T\left(P_Y \bar{B}_u +\bar{N}\right) +  \tilde{\bar{u}}^T \bar{M}  + \mu \left( 2\tilde{\bar{u}}^T W  + \bar{x}^T \bar{C}_v^T \right)   
\end{equation} 
Next, we show that Slater's condition \cite{slater2013lagrange} holds. Let $\bar{v}_i$ denote a nonzero component of the vector $\bar{C}_v\bar{x}$, let $W_{ii}$ denote the corresponding diagonal entry of matrix $W$, and choose $\tilde{\bar{u}}=-\left(\alpha v_i/W_{ii} \right) e_i$ where $\alpha$ is a real scalar and $e_i$ is a standard basis vector. Constraint \eqref{u_cond} is then reduced to 
\begin{equation}
    \left(\alpha^2 - \alpha \right)\left(\frac{v_i}{W_{ii}}\right)^2 \leq 0
\end{equation}
which clearly holds strictly for any $\alpha \in (0,1)$. As such, there always exists a $\tilde{\bar{u}}$ which is strictly feasible and hence Slater's condition is satisfied. 
Since optimization \eqref{of_controller2} is convex and Slater's condition holds, the KKT conditions are both necessary and sufficient for optimality. Thus any pair $(\tilde{\bar{u}},\mu)$ are primal and dual optimal if and only if the following conditions hold:
\begin{align}
\nabla_{\tilde{\bar{u}}}\mathcal{L} &= 0 \label{saddle_cond} \\
     \mu & \geq 0 \\
     \mu \left( \tilde{\bar{u}}^T W \tilde{\bar{u}} + \bar{x}^T \bar{C}_v^T \tilde{\bar{u}}\right) &=0  \label{slackness} \\
     \tilde{\bar{u}}^T W \tilde{\bar{u}} + \bar{x}^T \bar{C}_v^T \tilde{\bar{u}} &\leq 0 \label{u_cond}
\end{align}
Solving \eqref{saddle_cond} for $\tilde{\bar{u}}$ yields
\begin{equation} \label{u_sol}
    \tilde{\bar{u}} = -K_{PGC}\bar{x}
\end{equation}
where $K_{PGC} \triangleq \left(\bar{M}+2\mu W\right)^{-1} \left( \bar{B}_u^T P_Y + \bar{N}^T +\mu\bar{C}_v\right)$. Next, we note that constraint \eqref{u_cond} must always be active (i.e., $\tilde{\bar{u}}^T W \tilde{\bar{u}} + \tilde{\bar{u}}^T \bar{C}_v \bar{x} = 0$). If this were not true, then complementary slackness condition \eqref{slackness} would require that $\mu=0$. This would imply that the magnitude of some components of $\tilde{\bar{u}}$ could be made arbitrarily large (such that $\tilde{J}$ becomes arbitrarily small) without violating \eqref{saddle_cond}, as matrix $\bar{M}$ is singular. However, this is turn would clearly lead to a violation of \eqref{u_cond} since $W$ is positive definite. Accordingly, we substitute \eqref{u_sol} into $\tilde{\bar{u}}^T W \tilde{\bar{u}} + \tilde{\bar{u}}^T \bar{C}_v \bar{x} = 0$ and obtain
\begin{equation} \label{polynomial}
    \bar{x}^T K_{PGC}^T
     W K_{PGC}\bar{x} 
     - \bar{x}^T \bar{C}_v^T K_{PGC}\bar{x}  = 0
\end{equation}
Given a specific $\bar{x}$, the previous expression is a polynomial function of $\mu$ (since $K_{PGC}$ is a function of $\mu$). Because the KKT conditions are necessary, there is guaranteed to exist a positive real root of \eqref{polynomial} which corresponds to the optimal dual variable $\mu^\star$, which can be computed via a standard root-finding algorithm. This result is useful for real-time implementation of nonlinear feedback controller \eqref{of_controller2}, as solving for the roots of polynomial equations can be done extremely efficiently.

Specifically, in our HiL experiment, $\mu^\star$ was computed in real-time at a frequency of 500Hz on the dSpace DS1103 board using a bisection algorithm with a finite number of iterations. Then $\mu^\star$ was substituted into \eqref{u_sol} to obtain the optimal augmented control input $\bar{u}^\star$. We note that control law \eqref{of_controller2} requires knowledge of the full system state, including the disturbance model states. However, we assume that the base acceleration can be measured with negligible noise. As such it is possible to use an open-loop, reduced order observer to exactly obtain the disturbance states. This is explained subsequently. We first apply a coordinate transformation to $\mapp{W}$ to obtain 
\begin{equation}
    \tilde{A}_w = T A_w T^{-1} = \begin{bmatrix}
        0 & 1 \\ -\omega_w^2 & -2 \zeta_w \omega_w
    \end{bmatrix}
\end{equation}
\begin{equation}
    \tilde{B}_w = TB_w=\begin{bmatrix}
        2 \zeta_w \omega_w \sigma_w \\ 0
    \end{bmatrix}~,~~~
    \tilde{C}_w = C_w T^{-1} = \begin{bmatrix}
        1 & 0
    \end{bmatrix}
\end{equation}
where $T=\mathcal{O}(A_w,C_w)$ is the observability matrix associated with $\{A_w,C_w\}$ and the state vector in the new coordinates is given by $\tilde{x}_w = T x_w$. Clearly, if $w$ can be measured with negligible noise, then we have access to the state variable $\tilde{x}_{w,1}$ immediately given the form of $\tilde{C}_w$. It is straightforward to show (see e.g., \cite{luenberger1979introduction}) that the second state variable can be estimated using an observer of the form
\begin{equation}
    \tfrac{d}{dt}\hat{\tilde{x}}_{w,2} = -2 \zeta_w \omega_w \hat{\tilde{x}}_{w,2} -\omega_w^2 \tilde{x}_{w,1} 
\end{equation}
In addition, the estimation error $\tilde{x}_{w,2}-\hat{\tilde{x}}_{w,2}$ converges exponentially to 0 at a rate of $-2 \zeta_w \omega_w$, which in our case is equal to $2\pi$. The disturbance states in the original coordinate system are then obtained via the inverse transformation $x_w = T^{-1}\tilde{x}_w$.

Finally, we assume it is possible to measure the transducer force $f$ directly with negligible noise. We can then estimate the plant states exactly using a Luenberger observer \cite{luenberger1964observing} of the form
\begin{equation}
    \begin{array}{rl}
    \tfrac{d}{dt}\hat{x}_p =&\hspace{-5pt} A_b \hat{x}_p\hspace{-1pt} + \hspace{-1pt} B_{pf} f \hspace{-1pt}+\hspace{-1pt} B_{pw} w \hspace{-1pt}+\hspace{-1pt} L_p \left(v \hspace{-1pt}-\hspace{-1pt} \hat{v}\right) \\
                     \hat{v} =& \hspace{-5pt} C_{pv}\hat{x}_p
\end{array}
\end{equation}
where
\begin{equation}
    A_b \triangleq \begin{bmatrix}
        \bar{0} & I \\
        -M_b^{-1}K_b & -M_b^{-1}C_b
    \end{bmatrix},~~~ B_{pf} \triangleq \begin{bmatrix} \bar{0} \\ M_b^{-1}\Gamma_f   
    \end{bmatrix}
\end{equation}
and the observer gain $L_p$ is designed to obtain a desired error convergence rate.

It is not possible to analytically quantify the performance improvement provided by the nonlinear self-powered controller. Thus we only can only determine $J$ in this case via HiL experiment or numerical simulation.

\section{HiL Experimental Setup}
\label{sec:testbed}

\begin{figure}
    \centering
   \includegraphics[scale=0.495,trim={0.1cm 0cm 0cm 0cm},clip]{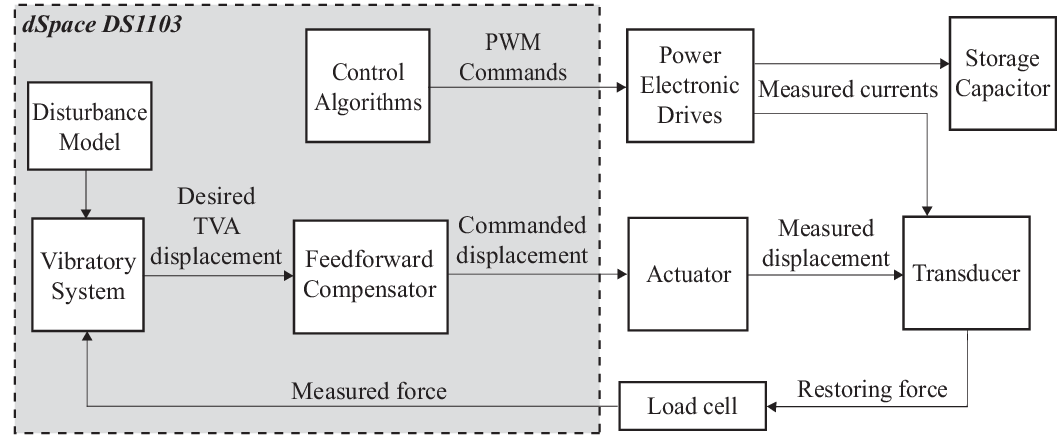}
   \caption{Simplified HiL testing block diagram}
   \label{fig:fig_hil}
\end{figure}

All of the three controllers described above were implemented and validated via HiL testing. HiL testing is an experimental method that interfaces numerical models with physical system components in real time. In the civil engineering literature HiL testing is known as real-time hybrid simulation \cite{blakeborough2001development}, and has been used extensively to study the performance of both structural control devices (e.g., \cite{christenson2008large,friedman2015large}) and vibratory energy harvesting technologies (e.g., \cite{cassidy2011design,asai2021hardware,ligeikis2024multiobjective}). A simplified block diagram of the HiL scheme used in this research is shown in Figure \ref{fig:fig_hil}. 

The plant dynamics given in \eqref{structure_dynamics}, and all dynamic systems associated with the control laws (i.e., $\mapp{Y},~\mapp{G},~\mapp{W}$, and the observers) were discretized using using Tustin's method. The discrete-time dynamics were simulated in Simulink on a dSpace DS1103 rapid prototyping board at a time step of $0.5$ms. The control algorithms for the power-electronic drives were executed at a faster rate of 10kHz, with measurement sampling synchronized with PWM switching to prevent signal corruption. Each HiL experiment lasted 10 minutes and data was recorded every 0.2ms. 

\begin{figure}
    \centering
   \includegraphics[scale=0.275]{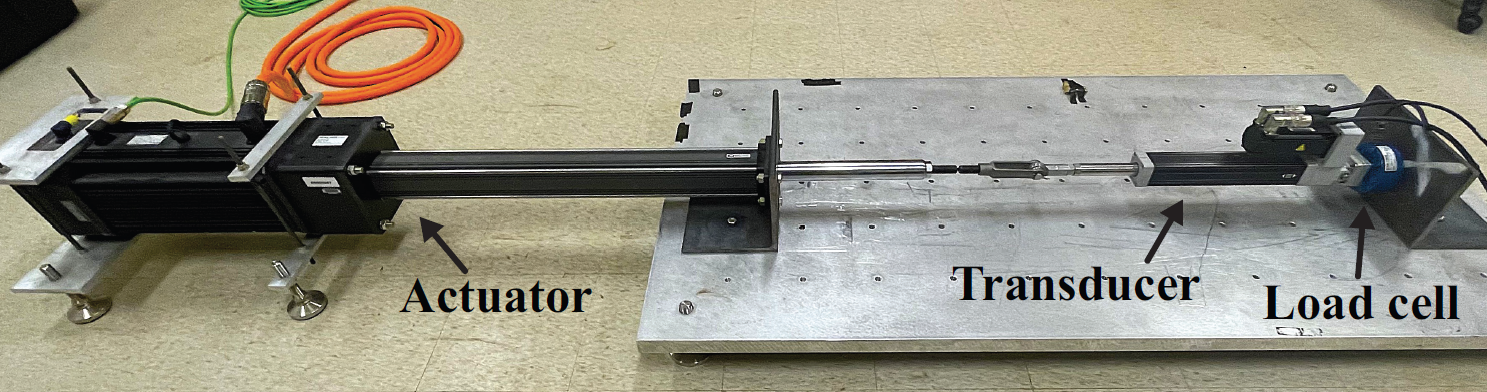}
   \caption{Experimental testbed}
   \label{fig:fig_exp}
\end{figure}

The physical testbed is shown in Figure \ref{fig:fig_exp}. It is comprised of a 50cm stroke, 30kN electromechanical linear actuator, which consists of a Exlar planetary roller screw coupled to a 20kW Lenze induction motor. The actuator position is controlled using a Lenze drive, which is interfaced with the dSpace DS1103 unit via the CAN protocol. The drive controller is highly configurable, with nested position, velocity, and current feedback loops. In addition, we employ a model-based, feed-forward compensator \cite{carrion2007model} to further reduce position-tracking errors. In this approach, the desired actuator displacement is passed through a pre-compensator which approximates an inverse model of the actuator dynamics. The compensated command is then sent to the Lenze drive. The transducer is attached to the actuator via a clevis connection, as shown in Figure \ref{fig:fig_exp}. An Interface Model 1210 load cell is used to measure the transducer force, and  an Analog Devices AD2S1205 resolver-to-digital converter chip is used to interface the transducer position/velocity measurements with the dSpace DS1103 board.

\section{Results}
\label{sec:rths_results}

\setlength{\tabcolsep}{4pt}
\begin{table}%[!htb]
\caption{Closed-loop performance $J$ results (with percent improvement relative to optimal static damping provided in parentheses)}
\small
\label{table:perf_data}
\begin{tabular}{l |c|c|c} 
\hline \vspace{1pt}
 Case  & Opt. damp.  & LTI SPSA & Nonlin. PGC \\ 
\hline\hline
Analytical  & 0.0243  & 0.0200 (17.7$\%$) &  --\\ 
HiL     & 0.0243  & 0.0204 (16.0$\%$) & 0.0191 (21.4$\%$) \\
Validation sim. & 0.0237 & 0.0202 (14.8$\%$)  & 0.0190 (19.8$\%$)   \\
Long sim. &0.0240  & 0.0198 (17.5$\%$)  & 0.0187 (22.1$\%$)   \\ 
\hline
\end{tabular}
\end{table}

\begin{table}
\caption{Closed-loop mean-square mass $m_1$ abs. acceleration $\Ex\{z_1^2\}$ in units of $(m/s^2)^2$ (with percent improvement relative to optimal static damping provided in parentheses)}
\small
\label{table:abs1_data}
\begin{tabular}{l |c|c|c} 
\hline \vspace{1pt}
 Case  & Opt. damp.  & LTI SPSA & Nonlin. PGC \\ 
\hline\hline
Analytical &  0.00715  & 0.00587 (17.9$\%$) &  --\\ 
HiL &  0.00713  &  0.00600 (15.8$\%$) &  0.00544 (23.7$\%$) \\
Validation sim. & 0.00703 &  0.00598 (14.9$\%$)  &  0.00543 (22.8$\%$)   \\
Long sim. & 0.00709  & 0.00584 (17.6$\%$)  & 0.00533 (24.8$\%$) \\ 
\hline
\end{tabular}
\end{table}

\begin{table}
\caption{Closed-loop mean-square mass $m_2$ abs. acceleration $\Ex\{z_2^2\}$ in units of $(m/s^2)^2$ (with percent improvement relative to optimal static damping provided in parentheses)}
\small
\label{table:abs2_data}
\begin{tabular}{l |c|c|c} 
\hline \vspace{1pt}
 Case  & Opt. damp.  & LTI SPSA & Nonlin. PGC \\ 
\hline\hline
Analytical &  0.0170  & 0.0138 (18.8$\%$) &  --\\ 
HiL &  0.0170  &  0.0141 (17.1$\%$) &  0.0133 (21.8$\%$) \\
Validation sim. & 0.0166 &0.0140 (15.7$\%$)  & 0.0132 (20.5$\%$)  \\
Long sim. & 0.0167  & 0.0137 (18.0$\%$) & 0.0130 (22.2$\%$)  \\ 
\hline
\end{tabular}
\end{table}

\begin{table}
\caption{Closed-loop mean-square transducer current $\Ex\{u^2\}$ in units of $A^2$ (with percent improvement relative to optimal static damping provided in parentheses)}
\small
\label{table:u_data}
\begin{tabular}{l |c|c|c} 
\hline \vspace{1pt}
 Case  & Opt. damp.  & LTI SPSA & Nonlin. PGC \\ 
\hline\hline
Analytical &  0.140 & 0.230 (-64.3$\%$) &  --\\ 
HiL &  0.131  &  0.228 (-74.0$\%$) &  0.269 (-105$\%$) \\
Validation sim.& 0.137 &  0.224 (-63.5$\%$) & 0.263 (-92.0$\%$)\\
Long sim. & 0.139  & 0.219 (-57.6$\%$) & 0.259 (-86.3$\%$) \\
\hline
\end{tabular}
\end{table}

Controller performance results are collected in Table \ref{table:perf_data}. The performance estimates obtained via HiL matched very closely with the analytically predicted performance for the optimal static damping and SPSA controllers. In addition, the nonlinear PGC controller provided more than a $21\%$ performance improvement over optimal static damping. 

\begin{figure}
    \centering
   \includegraphics[scale=0.55]{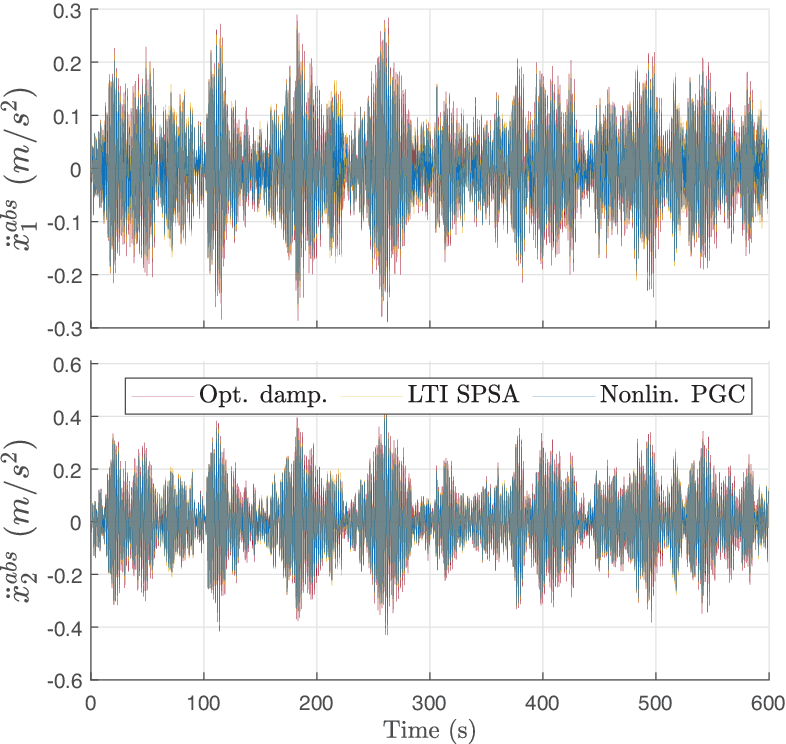} 
   \caption{Absolute acceleration data obtained from HiL experiments: mass $m_1$ (top) and mass $m_2$ (bottom)}
   \label{fig:rths_accel}
\end{figure}

\begin{figure}
    \centering
   \includegraphics[scale=0.54] {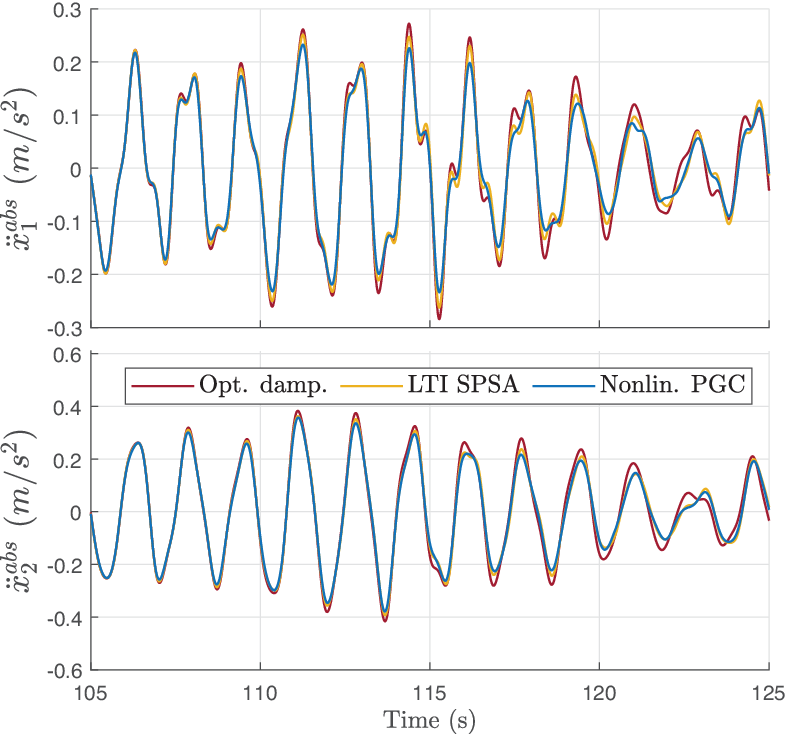} 
   \vspace{-5pt}
   \caption{Representative portion of absolute acceleration data obtained from HiL experiments: mass $m_1$ (top) and mass $m_2$ (bottom)}
   \label{fig:rths_accel_portion}
\end{figure}

\begin{figure}
    \centering
   \includegraphics[scale=0.55]{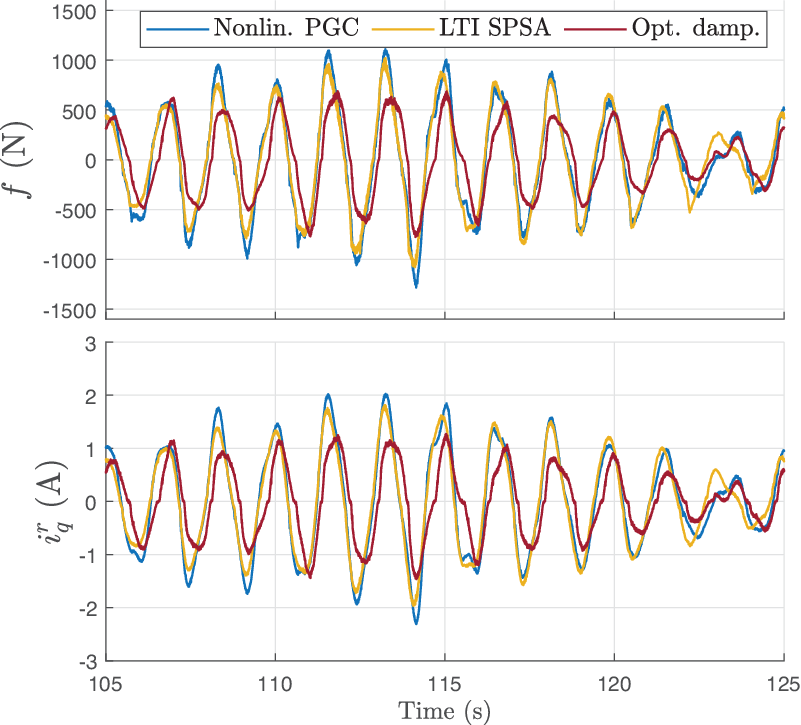} 
   \vspace{-5pt}
   \caption{Representative portion of transducer force (top) and quadrature-axis current (bottom) data obtained from HiL experiments}
   \label{fig:rths_iq_portion}
\end{figure}

Tables \ref{table:abs1_data} and \ref{table:abs2_data} report data on the mean-square absolute accelerations with the different control laws imposed. We obtain almost a 24$\%$ improvement in the mean-square acceleration of $m_1$ and a 22$\%$ improvement in the mean-square acceleration of $m_2$, with the nonlinear PGC controller. This reduction in accelerations is also apparent in Figures \ref{fig:rths_accel} and \ref{fig:rths_accel_portion}. Conversely, the mean-square transducer currents significantly increased for the SPSA and nonlinear PGC controllers, as indicated in Table \ref{table:u_data}. This is also evident in Figure \ref{fig:rths_iq_portion}. 

\begin{figure}
    \centering
   \includegraphics[scale=0.55]{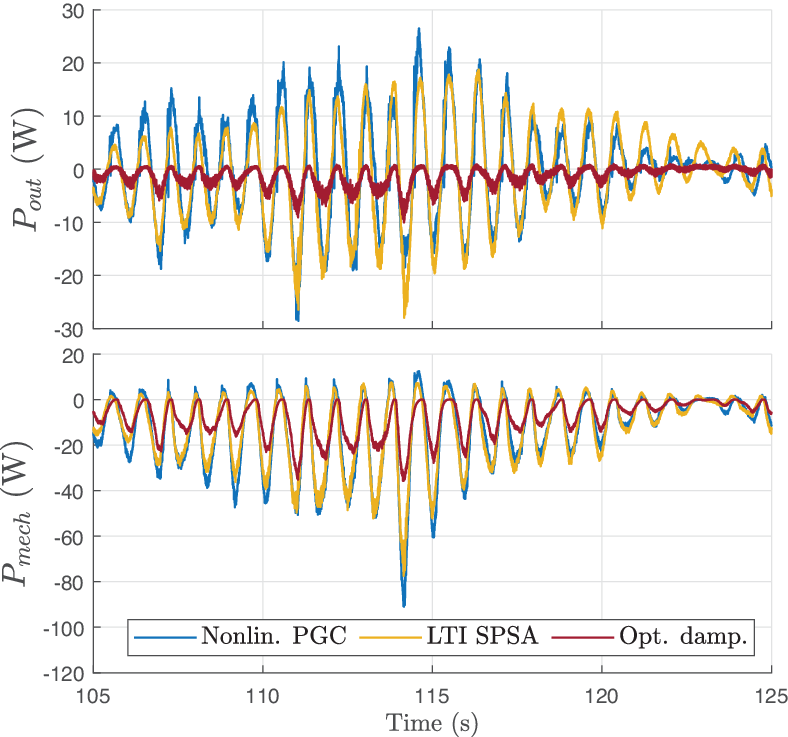} 
   \vspace{-5pt}
   \caption{Representative portion of measured DC-DC converter output electrical power (top) and transducer mechanical power (bottom) from each HiL experiment}
   \label{fig:rths_power_portion}
\end{figure}

\begin{figure}
    \centering
   \includegraphics[scale=0.55,trim={0.3cm 16.5cm 1cm 1.7cm},clip]{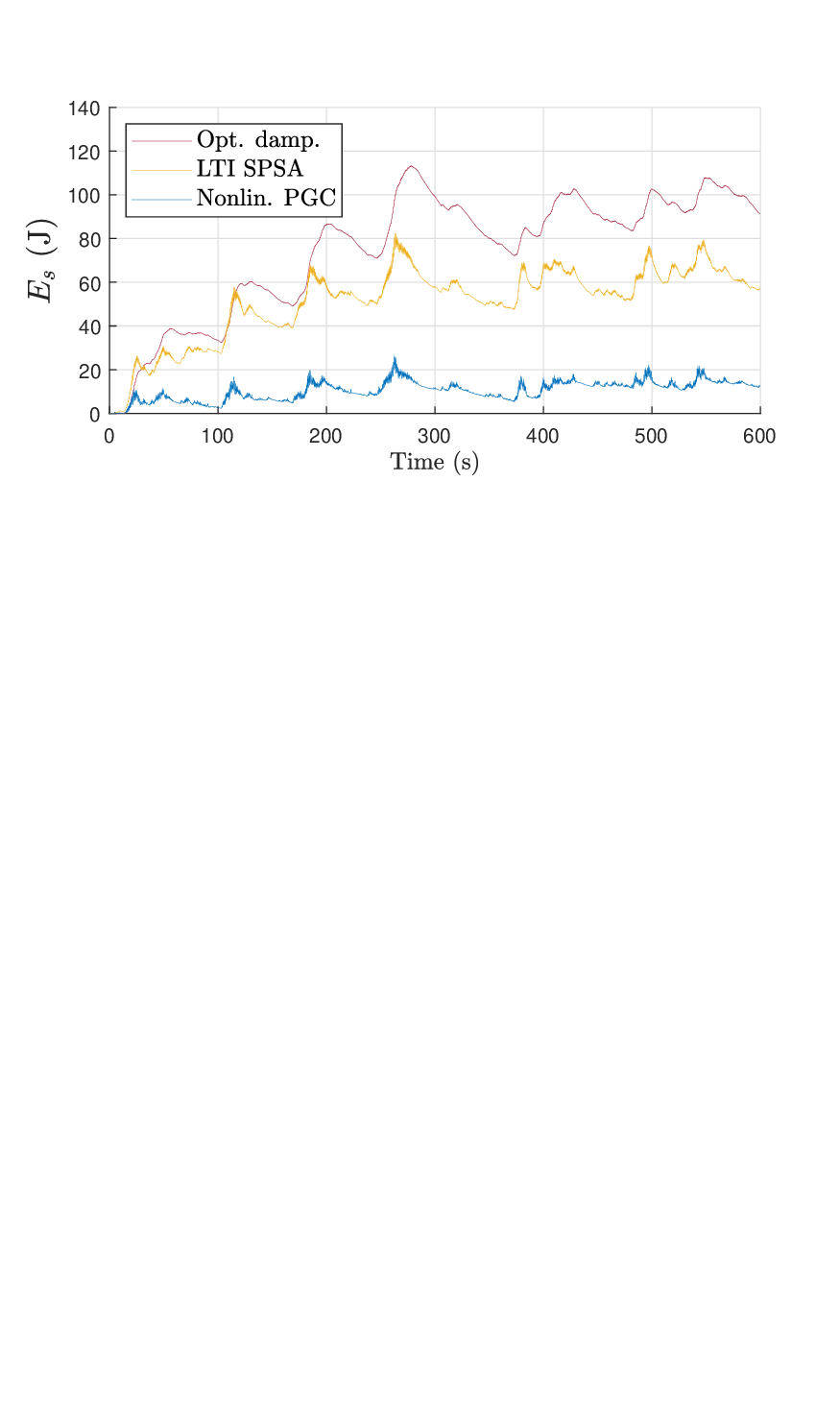}
   \caption{Evolution of stored energy during each HiL experiment}
   \label{fig:rths_energy}
\end{figure}

One explanation for the improved performance provided by the SPSA and nonlinear PGC controllers is that they allow for bidirectional electrical power flows, while the static damping controller obviously does not. This is most clearly illustrated in Figure \ref{fig:rths_power_portion}. These plots show the DC-DC converter output power, where negative $P_{out}$ implies power flowing into the storage capacitor. We see significantly larger power flows going from static damping to the SPSA, and again from the SPSA to the nonlinear controller. Furthermore, Figure \ref{fig:rths_power_portion} shows that there are occasional positive \emph{mechanical} transducer power flows, implying energy injection into the plant. We note that $P_{mech}$ was calculated as $P_{mech} = f \dot{x}_t$.
We also observe that there was significantly less energy accumulated in the capacitor with the nonlinear PGC controller imposed than in the other two cases, as shown in Figure \ref{fig:rths_energy}. It could be argued that this controller makes better use of the recycled energy.

\begin{figure}
    \centering
   \includegraphics[scale=0.55]{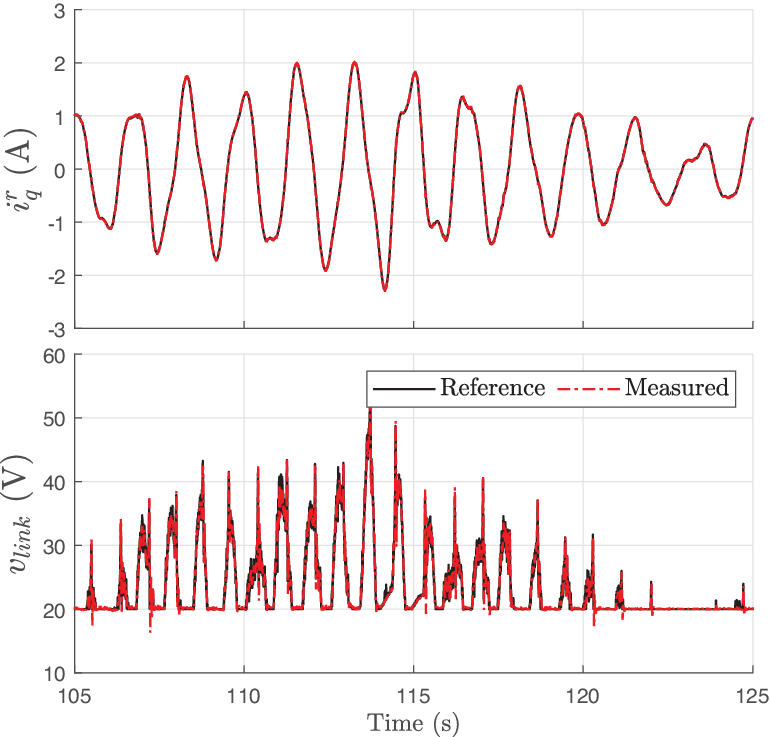} 
   \caption{Comparison of reference and measured quadrature-axis current signals (top) and DC link voltage signals (bottom) during HiL experiment with nonlinear PGC controller}
   \label{fig:rths_tracking}
\end{figure}

As shown in Figure \ref{fig:rths_tracking}, good tracking was maintained between the reference and measured $i_q^r$ and $v_{link}$ signals, indicating that the power-electronic control design approach worked as desired.

\begin{figure}
    \vspace{10pt}
    \centering
   \includegraphics[scale=0.554] {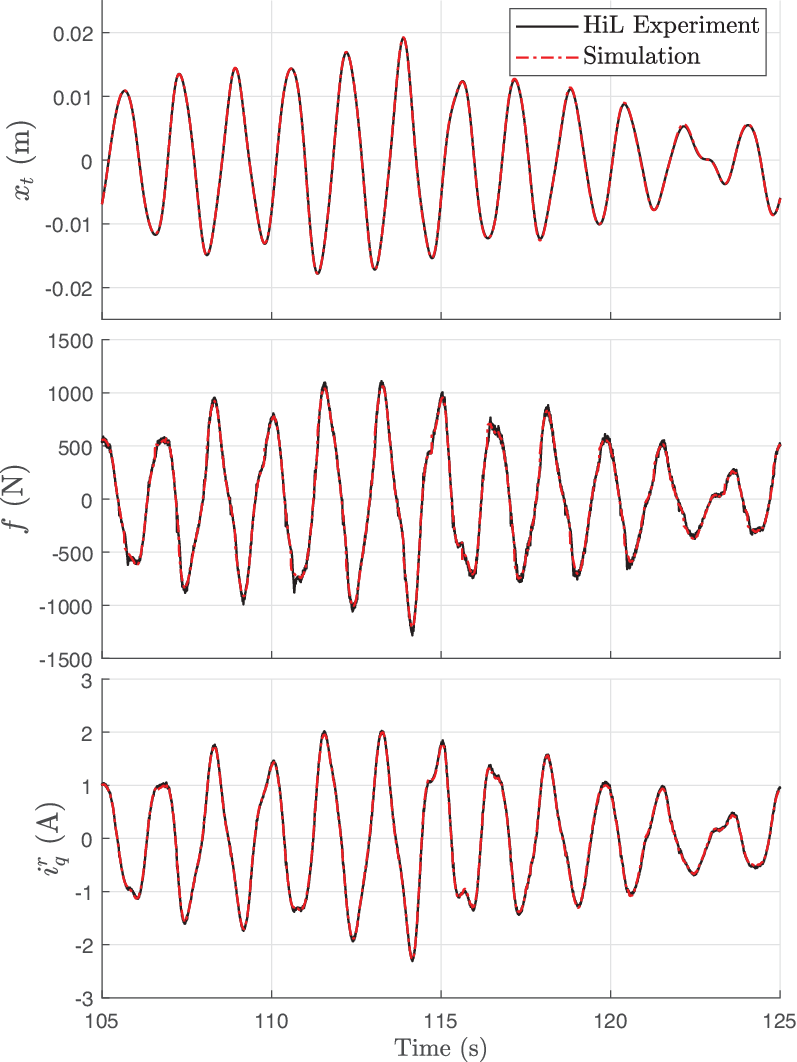} 
   \caption{Representative portion of transducer displacement (top); force (middle); and quadrature-axis current (bottom) data from HiL experiment and corresponding numerical simulation for nonlinear PGC controller}
   \label{fig:rths_sim_pgc_1}
\end{figure}

Finally, we conducted fully numerical simulations to validate the HiL results and our transducer model \eqref{trans_model}. We simulated the dynamic response of the full nonlinear system \eqref{structure_dynamics_3} in Simulink using the same 600s disturbance time history from the HiL experiments. We assumed instantaneous tracking of the current commands produced by the self-powered control laws (i.e., we did not simulate the DC link dynamics, the PWM switching of the power electronic drives, nor the dynamics of the low-level PI tracking loops). Figure \ref{fig:rths_sim_pgc_1} contains a variety of data comparing the HiL with the simulation results corresponding to the nonlinear self-powered feedback controller over a representative time-span. There was consistently very good agreement between all signals, confirming the validity of using \eqref{trans_model} to model the transducer’s mechanical dynamics. Although not shown here, we also found excellent agreement between the experimental and simulated time histories for the optimal static damping and SPSA controllers. We subsequently performed additional simulations for the much longer time span of 6000s to obtain more accurate performance estimates. The performance data obtained from these simulations are reported in Tables \ref{table:perf_data} through \ref{table:u_data}.

\section{Conclusion}
\label{sec:conclusion}

This article has detailed the design, construction, and experimental validation of a prototype self-powered vibration control system. We designed both linear and nonlinear self-powered feedback laws using a theoretical framework previously developed by the authors. The experimental results validated this framework and the functionality of our prototype system.

There is much future research to be done in the area of self-powered systems. For example, the theory utilized herein assumes a rather simplistic parasitic loss model and it could be extended to accommodate more realistic loss models (e.g., accounting for apparent static power loss). The methodology used for the DC-DC power converter controller design was heuristic, and while it performed well, a more systematic approach incorporating robust or nonlinear control techniques could be developed.
Finally, our HiL experimental validation only involved one transducer. Future experiments could be conducted in which there are several transducers connected to a common energy storage system. Such a demonstration of power-sharing between transducers would fully illustrate the capabilities of self-powered vibration control systems.

\bibliographystyle{IEEEtran}
\bibliography{main}

\begin{thebibliography}{10}
\providecommand{\url}[1]{#1}
\csname url@samestyle\endcsname
\providecommand{\newblock}{\relax}
\providecommand{\bibinfo}[2]{#2}
\providecommand{\BIBentrySTDinterwordspacing}{\spaceskip=0pt\relax}
\providecommand{\BIBentryALTinterwordstretchfactor}{4}
\providecommand{\BIBentryALTinterwordspacing}{\spaceskip=\fontdimen2\font plus
\BIBentryALTinterwordstretchfactor\fontdimen3\font minus \fontdimen4\font\relax}
\providecommand{\BIBforeignlanguage}[2]{{%
\expandafter\ifx\csname l@#1\endcsname\relax
\typeout{** WARNING: IEEEtran.bst: No hyphenation pattern has been}%
\typeout{** loaded for the language `#1'. Using the pattern for}%
\typeout{** the default language instead.}%
\else
\language=\csname l@#1\endcsname
\fi
#2}}
\providecommand{\BIBdecl}{\relax}
\BIBdecl

\bibitem{nakano2003self}
K.~Nakano, Y.~Suda, and S.~Nakadai, ``Self-powered active vibration control using a single electric actuator,'' \emph{Journal of Sound and Vibration}, vol. 260, no.~2, pp. 213--235, 2003.

\bibitem{nakano2004combined}
K.~Nakano, ``Combined type self-powered active vibration control of truck cabins,'' \emph{Vehicle System Dynamics}, vol.~41, no.~6, pp. 449--473, 2004.

\bibitem{khoshnoud2015energy}
F.~Khoshnoud, Y.~Zhang, R.~Shimura, A.~Shahba, G.~Jin, G.~Pissanidis, Y.~Chen, and C.~De~Silva, ``Energy regeneration from suspension dynamic modes and self-powered actuation,'' \emph{IEEE/ASME Transactions on Mechatronics}, vol.~20, no.~5, pp. 2513--2524, 2015.

\bibitem{choi2009self}
Y.~Choi and N.~Wereley, ``Self-powered magnetorheological dampers,'' \emph{Journal of Vibration and Acoustics}, vol. 131, no.~4, 2009.

\bibitem{tang2011self}
X.~Tang and L.~Zuo, ``Self-powered active control of structures with {TMDs},'' in \emph{Structural Dynamics and Renewable Energy, Volume 1}.\hskip 1em plus 0.5em minus 0.4em\relax Springer, 2011, pp. 227--238.

\bibitem{asai2016nonlinear}
T.~Asai and J.~Scruggs, ``Nonlinear stochastic control of self-powered variable-damping vibration control systems,'' in \emph{2016 American Control Conference (ACC)}, 2016, pp. 442--448.

\bibitem{li2022self}
J.-Y. Li and S.~Zhu, ``Self-powered active vibration control: concept, modeling, and testing,'' \emph{Engineering}, vol.~11, pp. 126--137, 2022.

\bibitem{cheng2024novel}
K.~Cheng, Y.~Guo, Z.~Li, Q.~Liang, L.~Li, and X.~Zhang, ``A novel self-powered pmsm electromagnetic damper based on adaptive economic model predictive control,'' \emph{IEEE Transactions on Transportation Electrification}, 2024.

\bibitem{jolly1997regenerative}
M.~Jolly and D.~Margolis, ``Regenerative systems for vibration control,'' \emph{Journal of Vibration and Acoustics}, vol. 119, pp. 208--215, 1997.

\bibitem{jolly1997assessing}
------, ``{Assessing the Potential for Energy Regeneration in Dynamic Subsystems},'' \emph{Journal of Dynamic Systems, Measurement, and Control}, vol. 119, no.~2, pp. 265--270, 06 1997.

\bibitem{margolis2005energy}
D.~Margolis, ``Energy regenerative actuator for motion control with application to fluid power systems,'' \emph{Journal of Dynamic Systems, Measurement, and Control}, vol. 127, pp. 33--40, 2005.

\bibitem{clemen2014model}
L.~Clemen, O.~Anubi, and D.~Margolis, ``Model predictive control of regenerative dampers with acceleration and energy harvesting trade-offs,'' in \emph{12th International Symposium on Advanced Vehicle Control, Tokyo, Japan, Sept}, 2014, pp. 22--26.

\bibitem{anubi2015energy}
O.~Anubi and L.~Clemen, ``Energy-regenerative model predictive control,'' \emph{Journal of the Franklin Institute}, vol. 352, no.~5, pp. 2152--2170, 2015.

\bibitem{clemen2016regenerative}
L.~Clemen, O.~Anubi, and D.~Margolis, ``On the regenerative capabilities of electrodynamic dampers using bond graphs and model predictive control,'' \emph{Journal of Dynamic Systems, Measurement, and Control}, vol. 138, no.~5, 2016.

\bibitem{liu2013regenerative}
Y.~Liu, L.~Zuo, and X.~Tang, ``Regenerative vibration control of tall buildings using model predictive control,'' in \emph{ASME Dynamic Systems and Control Conference}, vol. 56123, 2013, p. V001T15A012.

\bibitem{shen2018energy}
W.~Shen, S.~Zhu, Y.~Xu, and H.~Zhu, ``Energy regenerative tuned mass dampers in high-rise buildings,'' \emph{Structural Control and Health Monitoring}, vol.~25, no.~2, p. e2072, 2018.

\bibitem{onoda2003energy}
J.~Onoda, K.~Makihara, and K.~Minesugi, ``Energy-recycling semi-active method for vibration suppression with piezoelectric transducers,'' \emph{AIAA journal}, vol.~41, no.~4, pp. 711--719, 2003.

\bibitem{onoda2008performance}
J.~Onoda and K.~Makihara, ``Performance of simple and sophisticated control in energy-recycling semi-active vibration suppression,'' \emph{Journal of Vibration and Control}, vol.~14, no.~3, pp. 417--436, 2008.

\bibitem{ligeikis2023feasibility}
C.~H. Ligeikis and J.~T. Scruggs, ``On the feasibility of self-powered linear feedback control,'' \emph{IEEE Transactions on Automatic Control}, 2023.

\bibitem{ligeikis2021nonlinear}
C.~Ligeikis and J.~Scruggs, ``Nonlinear feedback controllers for self-powered systems with non-ideal energy storage subsystems,'' in \emph{2021 American Control Conference (ACC)}.\hskip 1em plus 0.5em minus 0.4em\relax IEEE, 2021, pp. 1748--1753.

\bibitem{ligeikis2021feasibility}
------, ``Feasibility and synthesis of finite-dimensional, linear time-invariant synthetic admittances for self-powered systems,'' in \emph{2021 IEEE 60th Conference on Decision and Control (CDC)}.\hskip 1em plus 0.5em minus 0.4em\relax IEEE, 2021, pp. 2440--2447.

\bibitem{ligeikis2022lqg}
------, ``An lqg-inspired framework for self-powered feedback control,'' in \emph{2022 IEEE 61st Conference on Decision and Control (CDC)}.\hskip 1em plus 0.5em minus 0.4em\relax IEEE, 2022, pp. 4519--4526.

\bibitem{CassidySPIE2011}
I.~Cassidy, J.~Scruggs, and S.~Behrens, ``Design of electromagnetic energy harvesters for large-scale structural vibration applications,'' \emph{SPIE Smart Structures and Materials/NDE}, pp. 1--11, 2011.

\bibitem{cassidy2012statistically}
I.~L. Cassidy and J.~T. Scruggs, ``Statistically linearized optimal control of an electromagnetic vibratory energy harvester,'' \emph{Smart Materials and Structures}, vol.~21, no.~8, p. 085003, 2012.

\bibitem{ligeikis2023self}
C.~Ligeikis, ``Self-powered systems for structural control: Theory and experiment,'' Ph.D. dissertation, University of Michigan, 2023.

\bibitem{krishnan2017permanent}
R.~Krishnan, \emph{Permanent magnet synchronous and brushless DC motor drives}.\hskip 1em plus 0.5em minus 0.4em\relax CRC press, 2017.

\bibitem{erickson2007fundamentals}
R.~W. Erickson and D.~Maksimovic, \emph{Fundamentals of power electronics}.\hskip 1em plus 0.5em minus 0.4em\relax Springer Science \& Business Media, 2007.

\bibitem{roberts2003random}
J.~B. Roberts and P.~D. Spanos, \emph{Random vibration and statistical linearization}.\hskip 1em plus 0.5em minus 0.4em\relax Courier Corporation, 2003.

\bibitem{scruggs2007non}
J.~Scruggs, A.~Taflanidis, and W.~Iwan, ``Non-linear stochastic controllers for semiactive and regenerative systems with guaranteed quadratic performance bounds—part 1: State feedback control,'' \emph{Structural Control and Health Monitoring}, vol.~14, no.~8, pp. 1101--1120, 2007.

\bibitem{scruggs2007non2}
------, ``Non-linear stochastic controllers for semiactive and regenerative systems with guaranteed quadratic performance bounds—part 2: output feedback control,'' \emph{Structural Control and Health Monitoring}, vol.~14, no.~8, p. 1121–1137, 2007.

\bibitem{scruggs2010multi}
J.~Scruggs, ``Multi-objective nonlinear control of semiactive and regenerative systems,'' in \emph{Proceedings of the American Control Conference}.\hskip 1em plus 0.5em minus 0.4em\relax IEEE, 2010, pp. 726--731.

\bibitem{cassidy2013nonlinear}
I.~L. Cassidy and J.~T. Scruggs, ``Nonlinear stochastic controllers for power-flow-constrained vibratory energy harvesters,'' \emph{Journal of Sound and Vibration}, vol. 332, no.~13, pp. 3134--3147, 2013.

\bibitem{slater2013lagrange}
M.~Slater, ``Lagrange multipliers revisited,'' in \emph{Traces and emergence of nonlinear programming}.\hskip 1em plus 0.5em minus 0.4em\relax Springer, 2013, pp. 293--306.

\bibitem{luenberger1979introduction}
D.~G. Luenberger, \emph{Introduction to dynamic systems: theory, models, and applications}.\hskip 1em plus 0.5em minus 0.4em\relax Wiley New York, 1979, vol.~1.

\bibitem{luenberger1964observing}
------, ``Observing the state of a linear system,'' \emph{IEEE transactions on military electronics}, vol.~8, no.~2, pp. 74--80, 1964.

\bibitem{blakeborough2001development}
A.~Blakeborough, M.~Williams, A.~Darby, and D.~Williams, ``The development of real--time substructure testing,'' \emph{Philosophical Transactions of the Royal Society of London. Series A: Mathematical, Physical and Engineering Sciences}, vol. 359, no. 1786, pp. 1869--1891, 2001.

\bibitem{christenson2008large}
R.~Christenson, Y.~Z. Lin, A.~Emmons, and B.~Bass, ``Large-scale experimental verification of semiactive control through real-time hybrid simulation,'' \emph{Journal of Structural Engineering}, vol. 134, no.~4, pp. 522--534, 2008.

\bibitem{friedman2015large}
A.~Friedman, S.~J. Dyke, B.~Phillips, R.~Ahn, B.~Dong, Y.~Chae, N.~Castaneda, Z.~Jiang, J.~Zhang, Y.~Cha \emph{et~al.}, ``Large-scale real-time hybrid simulation for evaluation of advanced damping system performance,'' \emph{Journal of Structural Engineering}, vol. 141, no.~6, p. 04014150, 2015.

\bibitem{cassidy2011design}
I.~L. Cassidy, J.~T. Scruggs, S.~Behrens, and H.~P. Gavin, ``Design and experimental characterization of an electromagnetic transducer for large-scale vibratory energy harvesting applications,'' \emph{Journal of Intelligent Material Systems and Structures}, vol.~22, no.~17, pp. 2009--2024, 2011.

\bibitem{asai2021hardware}
T.~Asai, M.~Takino, Y.~Watanabe, and K.~Sugiura, ``Hardware-in-the-loop testing of an electromagnetic transducer with a tuned inerter for vibratory energy harvesting,'' \emph{ASCE-ASME J Risk and Uncert in Engrg Sys Part B Mech Engrg}, vol.~7, no.~1, 2021.

\bibitem{ligeikis2024multiobjective}
C.~H. Ligeikis and J.~T. Scruggs, ``Multiobjective vector control of a three-phase vibratory energy harvester,'' \emph{IEEE Transactions on Control Systems Technology}, vol.~32, no.~5, pp. 1770--1784, 2024.

\bibitem{carrion2007model}
J.~E. Carrion, \emph{Model-based strategies for real-time hybrid testing}.\hskip 1em plus 0.5em minus 0.4em\relax University of Illinois at Urbana-Champaign, 2007.

\end{thebibliography}

\end{document}